\input harvmac
%\draftmode
\noblackbox
%Ram

\let\includefigures=\iftrue
\let\useblackboard=\iftrue
\newfam\black

%Figure Stuff
\includefigures
\message{If you do not have epsf.tex (to include figures),}
\message{change the option at the top of the tex file.}
\input epsf
\def\figin{\epsfcheck\figin}\def\figins{\epsfcheck\figins}
\def\epsfcheck{\ifx\epsfbox\UnDeFiNeD
\message{(NO epsf.tex, FIGURES WILL BE IGNORED)}
\gdef\figin##1{\vskip2in}\gdef\figins##1{\hskip.5in}% blank space instead
\else\message{(FIGURES WILL BE INCLUDED)}%
\gdef\figin##1{##1}\gdef\figins##1{##1}\fi}
\def\DefWarn#1{}
\def\figinsert{\goodbreak\midinsert}
\def\ifig#1#2#3{\DefWarn#1\xdef#1{fig.~\the\figno}
\writedef{#1\leftbracket fig.\noexpand~\the\figno}%
\figinsert\figin{\centerline{#3}}\medskip\centerline{\vbox{
\baselineskip12pt\advance\hsize by -1truein
\noindent\footnotefont{\bf Fig.~\the\figno:} #2}}
\bigskip\endinsert\global\advance\figno by1}
%%%
\else
\def\ifig#1#2#3{\xdef#1{fig.~\the\figno}
\writedef{#1\leftbracket fig.\noexpand~\the\figno}%
%\figinsert\figin{\centerline{#3}}\medskip
%\centerline{\vbox{\baselineskip12pt
%\advance\hsize by -1truein\noindent
%\footnotefont{\bf Fig.~\the\figno:} #2}}
%\bigskip\endinsert
\global\advance\figno by1}
\fi

\def\doublefig#1#2#3#4{\DefWarn#1\xdef#1{fig.~\the\figno}
\writedef{#1\leftbracket fig.\noexpand~\the\figno}%
\figinsert\figin{\centerline{#3\hskip1.0cm#4}}\medskip\centerline{\vbox{
\baselineskip12pt\advance\hsize by -1truein
\noindent\footnotefont{\bf Fig.~\the\figno:} #2}}
\bigskip\endinsert\global\advance\figno by1}

\def\bar{\overline}
\def\CS{{\cal S}}
\def\CN{{\cal N}}
\def\CH{{\cal H}}

\def\rt2{\sqrt{2}}
\def\irt2{{1\over\sqrt{2}}}

\def\tt{\widetilde}

%\deMelloKochIR
\lref\deMelloKochIR{ R.~de Mello Koch and J.~P.~Rodrigues,
%``The Collective field theory of a singular supersymmetric matrix model,''
Phys.\ Rev.\ D {\bf 51}, 5847 (1995) [arXiv:hep-th/9410012].
%%CITATION = HEP-TH 9410012;%%
}

%\ChaudhuriYK
\lref\chaudpol{ S.~Chaudhuri and J.~Polchinski,
``Critical behavior of the Marinari-Parisi model,''
Phys.\ Lett.\ B {\bf 339}, 309 (1994) [arXiv:hep-th/9408005].
%%CITATION = HEP-TH 9408005;%%
}

\lref\MarinariJC{
E.~Marinari and G.~Parisi,
``The Supersymmetric One-Dimensional String,''
Phys.\ Lett.\ B {\bf 240}, 375 (1990).
%%CITATION = PHLTA,B240,375;%%
}

\lref\FerrettiFU{ G.~Ferretti,
``The Untruncated Marinari-Parisi Superstring,''
J.\ Math.\ Phys.\  {\bf 35}, 4469 (1994) [arXiv:hep-th/9310002].
%%CITATION = HEP-TH 9310002;%%
}

\lref\BrusteinNC{ R.~Brustein, M.~Faux and B.~A.~Ovrut,
``Effective d = 2 supersymmetric Lagrangians from d = 1 supermatrix models,''
Nucl.\ Phys.\ B {\bf 421}, 293 (1994) [arXiv:hep-th/9310033];
%%CITATION = HEP-TH 9310033;%%
R.~Brustein, M.~Faux and B.~A.~Ovrut,
``Supersymmetric field theory from supermatrix models,''
arXiv:hep-th/9407164.
%%CITATION = HEP-TH 9407164;%%
}

\lref\Feinberg{
J.~Feinberg,
``Dissipation of the string embedding dimension in the singlet sector of a
%matrix model at large alpha-prime,''
Nucl.\ Phys.\ B {\bf 405}, 389 (1993)
[arXiv:hep-th/9304059];
%%CITATION = HEP-TH 9304059;%%
J.~Feinberg,
``Stabilized matrix models for nonperturbative two-dimensional quantum
gravity,''
Int.\ J.\ Mod.\ Phys.\ A {\bf 9}, 3751 (1994)
[arXiv:hep-th/9212069];
%%CITATION = HEP-TH 9212069;%%
J.~Feinberg,
``String field theory for d $\leq$ 0
matrix models via Marinari-Parisi,''
Phys.\ Lett.\ B {\bf 281}, 225 (1992)
[arXiv:hep-th/9201037].
%%CITATION = HEP-TH 9201037;%%
}

%\KarlinerCD
\lref\KarlinerCD{
M.~Karliner and S.~Migdal,
``Nonperturbative 2-D Quantum Gravity Via Supersymmetric String,''
Mod.\ Phys.\ Lett.\ A {\bf 5}, 2565 (1990).
%%CITATION = MPLAE,A5,2565;%%
}

%\GonzalezYU
\lref\GonzalezYU{
J.~Gonzalez,
``A New Vacuum For The Supersymmetric One-Dimensional Discretized String,''
Phys.\ Lett.\ B {\bf 255}, 367 (1991);
%%CITATION = PHLTA,B255,367;%%
S.~Bellucci, T.~R.~Govindrajan, A.~Kumar and R.~N.~Oerter,
``On The Nonperturbative Solution Of D = 1 Superstring,''
Phys.\ Lett.\ B {\bf 249}, 49 (1990);
%%CITATION = PHLTA,B249,49;%%
S.~Bellucci,
``Duality Transformation Of The One-Dimensional Supersymmetric String,''
Phys.\ Lett.\ B {\bf 257}, 35 (1991);
%%CITATION = PHLTA,B257,35;%%
``Broken supersymmetry in the matrix model on a circle,''
Z.\ Phys.\ C {\bf 54}, 565 (1992).
%%CITATION = ZEPYA,C54,565;%
}

\lref\vanTonderVC{ A.~J.~van Tonder,
``A Continuum description of superCalogero models,''
arXiv:hep-th/9204034.
%%CITATION = HEP-TH 9204034;%%
}

\lref\RodriguesBY{ J.~P.~Rodrigues and A.~J.~van Tonder,
``Marinari-Parisi and supersymmetric collective field theory,''
Int.\ J.\ Mod.\ Phys.\ A {\bf 8}, 2517 (1993)
[arXiv:hep-th/9204061].
%%CITATION = HEP-TH 9204061;%%
}

\lref\JevickiYK{ A.~Jevicki and J.~P.~Rodrigues,
``Supersymmetric collective field theory,''
Phys.\ Lett.\ B {\bf 268}, 53 (1991).
%%CITATION = PHLTA,B268,53;%%
}

\lref\CohnZJ{
J.~D.~Cohn and H.~Dykstra,
``The Marinari-Parisi model and collective field theory,''
Mod.\ Phys.\ Lett.\ A {\bf 7}, 1163 (1992)
[arXiv:hep-th/9202050].
%%CITATION = HEP-TH 9202050;%%
}

\lref\ZZB{ A.~B.~Zamolodchikov and A.~B.~Zamolodchikov,
``Liouville field theory on a pseudosphere,''
arXiv:hep-th/0101152.
%%CITATION = HEP-TH 0101152;%%
}

%\MartinecKA
\lref\MartinecKA{ E.~J.~Martinec, ``The annular report on
non-critical string theory,'' arXiv:hep-th/0305148.
%%CITATION = HEP-TH 0305148;%%
}

%\KlebanovKM
\lref\KlebanovKM{ I.~R.~Klebanov, J.~Maldacena and N.~Seiberg,
``D-brane decay in two-dimensional string theory,''
arXiv:hep-th/0305159.
%%CITATION = HEP-TH 0305159;%%
}

%\McGreevyEP
\lref\McGreevyEP{ J.~McGreevy, J.~Teschner and H.~Verlinde,
``Classical and quantum D-branes in 2D string theory,''
arXiv:hep-th/0305194.
%%CITATION = HEP-TH 0305194;%%
}

%\McGreevyKB
\lref\McGreevyKB{ J.~McGreevy and H.~Verlinde, ``Strings from
tachyons: The c = 1 matrix reloaded,'' arXiv:hep-th/0304224.
%%CITATION = HEP-TH 0304224;%%
}

%\BoucherBH
\lref\BoucherBH{ W.~Boucher, D.~Friedan and A.~Kent, ``Determinant
Formulae And Unitarity For The N=2 Superconformal Algebras In
Two-Dimensions Or Exact Results On String Compactification,''
Phys.\ Lett.\ B {\bf 172}, 316 (1986).
%%CITATION = PHLTA,B172,316;%%
}

%\AhnSX
\lref\AhnSX{ C.~Ahn, C.~Kim, C.~Rim and M.~Stanishkov, ``Duality
in N = 2 super-Liouville theory,'' arXiv:hep-th/0210208.
%%CITATION = HEP-TH 0210208;%%
}

\lref\AhnEV{ C.~Ahn, C.~Rim and M.~Stanishkov, ``Exact one-point
function of N = 1 super-Liouville theory with boundary,'' Nucl.\
Phys.\ B {\bf 636}, 497 (2002) [arXiv:hep-th/0202043].
%%CITATION = HEP-TH 0202043;%%
}

\lref\FukudaBV{ T.~Fukuda and K.~Hosomichi, ``Super Liouville
theory with boundary,'' Nucl.\ Phys.\ B {\bf 635}, 215 (2002)
[arXiv:hep-th/0202032].
%%CITATION = HEP-TH 0202032;%%
}

\lref\FZZ{ V.~Fateev, A.~B.~Zamolodchikov and A.~B.~Zamolodchikov,
``Boundary Liouville field theory. I: Boundary state and boundary
\ two-point function,'' arXiv:hep-th/0001012.
%%CITATION = HEP-TH 0001012;%%
}

\lref\TLbound{ J.~Teschner, ``Remarks on Liouville theory with
boundary,'' arXiv:hep-th/0009138.
%%CITATION = HEP-TH 0009138;%%
}

%\ThompsonRW
\lref\ThompsonRW{ D.~M.~Thompson, ``Descent relations in type 0A /
0B,'' Phys.\ Rev.\ D {\bf 65}, 106005 (2002)
[arXiv:hep-th/0105314].
%%CITATION = HEP-TH 0105314;%%
}

%\IshibashiKG
\lref\IshibashiKG{ N.~Ishibashi, ``The Boundary And Crosscap
States In Conformal Field Theories,'' Mod.\ Phys.\ Lett.\ A {\bf
4}, 251 (1989).
%%CITATION = MPLAE,A4,251;%%
}

%\MartinecBG
\lref\MartinecBG{ E.~J.~Martinec and G.~M.~Sotkov, ``Superghosts
Revisited: Supersymmetric Bosonization,'' Phys.\ Lett.\ B {\bf
208}, 249 (1988).
%%CITATION = PHLTA,B208,249;%%
}

\lref\TakayanagiSM{
T.~Takayanagi and N.~Toumbas,
``A matrix model dual of type 0B string theory in two dimensions,''
arXiv:hep-th/0307083.
%%CITATION = HEP-TH 0307083;%%
}

\def\bra#1{{\langle}#1|}
\def\ket#1{|#1\rangle}
\def\bbra#1{{\langle\langle}#1|}
\def\kket#1{|#1\rangle\rangle}
\def\vev#1{\langle{#1}\rangle}

\def\CA{{\cal A}}

\def\CF{{\cal F}}

\def\CN{{\cal N}}
%AEL

\def\CL{{\cal L}}

\def\CS{{\cal S}}

%AEL

\def\IZ{{\bf Z}}
\def\subsubsec#1{\bigskip\noindent{\it{#1}}\bigskip}
\def\downvac{\ket{\downarrow\downarrow\cdots\downarrow}}

\def\ie{{\it i.e.}}
\def\eg{{\it e.g.}}
\def\qv{{\it q.v.}}
\def\cf{{\it c.f.}}
\def\IZ{{\bf Z}}
%\def\IZ{\relax{\rm Z\kern-.03em Z}}
%\relax\ifmmode\mathchoice
%{\hbox{\cmss Z\kern-.4em Z}}{\hbox{\cmss Z\kern-.4em Z}}
%{\lower.9pt\hbox{\cmsss Z\kern-.4em Z}}
%{\lower1.2pt\hbox{\cmsss Z\kern-.4em Z}}
%\else{\cmss Z\kern-.4emZ}\fi}
%{\cmss Z\kern-.4emZ}}
\def\IB{\relax{\rm I\kern-.18em B}}
\def\IC{{\relax\hbox{$\inbar\kern-.3em{\rm C}$}}}
\def\ID{\relax{\rm I\kern-.18em D}}
\def\IE{\relax{\rm I\kern-.18em E}}
\def\IF{\relax{\rm I\kern-.18em F}}
\def\IG{\relax\hbox{$\inbar\kern-.3em{\rm G}$}}
\def\IGa{\relax\hbox{${\rm I}\kern-.18em\Gamma$}}
\def\IH{\relax{\rm I\kern-.18em H}}
\def\II{\relax{\rm I\kern-.18em I}}
\def\IK{\relax{\rm I\kern-.18em K}}
\def\IP{\relax{\rm I\kern-.18em P}}
\def\IR{\relax{\rm I\kern-.18em R}}
\def\mfl{ (-1)^{F_L} }
\def\jhep#1#2#3{JHEP\ {\bf #1} #3 (19#2)}

%\DabholkarTE
\lref\atish{
A.~Dabholkar,
``Fermions and nonperturbative supersymmetry breaking in the one-dimensional superstring,''
Nucl.\ Phys.\ B {\bf 368}, 293 (1992).
%%CITATION = NUPHA,B368,293;%%
}

%\GaberdielJR
\lref\GaberdielJR{
M.~R.~Gaberdiel,
``Lectures on non-BPS Dirichlet branes,''
Class.\ Quant.\ Grav.\  {\bf 17}, 3483 (2000)
[arXiv:hep-th/0005029], and references therein.
%%CITATION = HEP-TH 0005029;%%
}

\lref\kutsei{D.~Kutasov and N.~Seiberg, ``Non-critical superstrings'', Phys.\ Lett.\ B {\bf 251}, 67 (1990).}

\lref\kutgivads{A.~Giveon and D.~Kutasov, ``Notes on $AdS_3$'', Nucl.\ Phys.\ B {\bf 621}, 303 (2002)  [arXiv:hep-th/9909110]. }

\lref\kutgivpelc{A.~Giveon, D.~Kutasov and O.~Pelc, ``Holography for noncritical superstrings'', JHEP {\bf 9910} 35 (1999) [arXiv:hep-th/9907178]. }

\lref\kutasov{D.~Kutasov, ``Some properties of (non)critical strings'', [arXiv:hep-th/9110041].}

\lref\wittenbh{E.~Witten, ``String theory and black holes'', Phys.\ Rev.\ D {\bf 44}, 314 (1991) \semi
G.~Mandal, A.~M.~Sengupta and S.~R.~Wadia, ``Classical solutions of two-dimensional string theory'', Mod.\ Phys.\ Lett. A {\bf 6}, 1685 (1991) \semi
S.~Elitzur, A.~Forge and E.~Rabinovici, ``Some global aspects of string compactifications'', Mod.\ Phys.\ Lett. A {\bf 6}, 1685 (1991).}

\lref\dvv{R.~Dijkgraaf, H.~Verlinde and E.~Verlinde, ``String propagation in a black hole geometry'',  Nucl.\ Phys.\ B {\bf 371}, 269 (1992).}

\lref\eguchi{T.~Eguchi and Y.Sugawara, ``Modular invariance in superstring on Calabi-Yau n-fold with A-D-E singularity'',  Nucl.\ Phys.\ B {\bf 577}, 3 (2000) [arXiv:hep-th/0002100].  }

\lref\mizo{S.~Mizoguchi, ``Modular invariant critical superstrings on four-dimensional Minkowski space $\times$ two-dimensional black hole'', JHEP {\bf 0004} 14 (2000) [arXiv:hep-th/0003053]. }

\lref\bilal{A.~Bilal and J.~L.~Gervais, ``New critical dimensions for string theories'', Nucl.\ Phys.\ B {\bf 284}, 397 (1987). }

\lref\farkas{H.~M.~Farkas and I.~Kra, ``Theta constants, Riemann surfaces and the modular group'', Graduate studies in mathematics, Vol.37. Amer.\ Math.\ Soc.\ }

\lref\horikap{K.~Hori and A.~Kapustin, ``Duality of the fermionic 2-d black hole and $\CN=2$ Liouville theory as mirror symmetry'', JHEP {\bf 0108} 045 (2001) [arXiv:hep-th/0104202].}

\lref\oogvafa{H.~Ooguri and C.~Vafa, ``Two-dimensional black hole and singularities of CY manifolds'', Nucl.\ Phys.\ B {\bf 463}, 55 (1996) [arXiv:hep-th/9511164]. }

\lref\dilholo{O.~Aharony, M.~Berkooz, D.~Kutasov and N.~Seiberg, ``Linear dilatons, NS5-branes and holography'', JHEP {\bf 9810} 004 (1998) [arXiv:hep-th/9808149]. }

\lref\seiberg{N.~Seiberg, ``Notes on quantum Liouville theory and quantum gravity'', Prog.\ Theor.\ Phys.\ Suppl {\bf 102} 319 (1990).}

\lref\maldoog{J.~Maldacena and H.~Ooguri,``Strings in $AdS_3$ and $SL_2(\IR)$ WZW model 1: The spectrum'', J.\ Math.\ Phys.\ {\bf 42} 2929 (2001)  [arXiv:hep-th/0001053]. }

\lref\fms{D.~Friedan, E.~Martinec and S.~Shenker, ``Conformal invariance, supersymmetry and string theory'', Nucl.\ Phys.\ B {\bf 271}, 93 (1986). }

\lref\chs{C.~Callan, J.~Harvey and A.~Strominger, ``Supersymmetric string solitons'', [arXiv:hep-th/9112030], in Trieste 1991, Proceedings, String theory and Quantum Gravity 1991, 208.}

\lref\polchinski{J.~Polchinski, ``String theory, Vol. I, II'', Cambridge University Press (1998).}

\lref\kleb{C.~P.~Herzog, I.~R.~Klebanov and P.~Ouyang, ``Remarks on the warped deformed conifold'',  [arXiv:hep-th/0108101].}

\lref\seiwitt{N.~Seiberg and E.~Witten, ``The $D1/D5$ system and singular CFT'', JHEP {\bf 9904} 017 (1999), [arxiv:hepth/9903224].}

\lref\mukhi{K.~Dasgupta and S.~Mukhi,``Brane constructions, conifolds and M-theory'', Nucl.\ Phys.\ B {\bf 551}, 204 (1999), [arxiv:hepth/9811139]. }

%\MurthyES
\lref\sameer{
S.~Murthy,
``Notes on non-critical superstrings in various dimensions,''
arXiv:hep-th/0305197.
%%CITATION = HEP-TH 0305197;%%
}

\lref\tong{D.~Tong, ``NS5-branes, T-duality and worldsheet instantons'', JHEP {\bf 0207} 013 (2002) [arXiv:hep-th/0204186].}

\lref\juan{J.~Maldacena, Private communication.}

\lref\kazsuz{Y.~Kazama and H.~Suzuki, ``New $N=2$ superconformal field theories and superstring compactification'', Nucl.\ Phys.\ B {\bf 321}, 232 (1989). }

\lref\harmoo{R.~Gregory, J.~Harvey and G.~Moore,``Unwinding strings and T-duality of Kaluza-Klein and H-monopoles'', Adv.\ Theor.\ Math.\ Phys.\ {\bf 1} 283 (1997), [arXiv:hep-th/9708086]. }

\lref\klemvafa{A.~Klemm, W.~Lerche, P.~Mayr, C.~Vafa and N.~Warner, ``Self-dual strings and $N=2$ supersymmetric field theory'', Nucl.\ Phys.\ B {\bf 477}, 746 (1996),  [arXiv:hep-th/9604034]\semi
S.~Gukov, C.~Vafa and E.~Witten, ``CFT's from Calabi-Yau four-folds'', Nucl.\ Phys.\ B {\bf 584}, 69 (2000),  [arXiv:hep-th/9906070]. }

\lref\polya{A.~M.~Polyakov, ``Quantum Geometry of Bosonic Strings'', Phys.\ Lett. B {\bf 103}, 207 (1981).}

%\BerkovitsTG
\lref\BerkovitsTG{
N.~Berkovits, S.~Gukov and B.~C.~Vallilo,
``Superstrings in 2D backgrounds with R-R flux and new extremal black  holes,''
Nucl.\ Phys.\ B {\bf 614}, 195 (2001)
[arXiv:hep-th/0107140].
%%CITATION = HEP-TH 0107140;%%
}

\lref\KS{
D.~Kutasov and N.~Seiberg,
``Noncritical Superstrings,''
Phys.\ Lett.\ B {\bf 251}, 67 (1990).
%%CITATION = PHLTA,B251,67;%%
}

\lref\threeK{
V.~Kazakov, I.~K.~Kostov and D.~Kutasov,
``A matrix model for the two-dimensional black hole,''
Nucl.\ Phys.\ B {\bf 622}, 141 (2002)
[arXiv:hep-th/0101011].
%%CITATION = HEP-TH 0101011;%%
}

\lref\BIPZ{
E.~Brezin, C.~Itzykson, G.~Parisi and J.~B.~Zuber,
``Planar Diagrams,''
Commun.\ Math.\ Phys.\  {\bf 59}, 35 (1978).
%%CITATION = CMPHA,59,35;%%
}

%\BaiET
\lref\BaiET{
J.~Z.~Bai {\it et al.}  [BES Collaboration],
``Measurements of the mass and full-width of the eta/c meson,''
Phys.\ Lett.\ B {\bf 555}, 174 (2003)
[arXiv:hep-ex/0301004].
%%CITATION = HEP-EX 0301004;%%
}

\lref\HoriAX{
V.~A.~Fateev, A.~B.~Zamolodchikov and Al.~B.~Zamolodchikov, unpublished;
A.~Giveon and D.~Kutasov, ``Little string theory in a double scaling limit'', JHEP {\bf 9910} 34 (1999) [arXiv:hep-th/9909110];
K.~Hori and A.~Kapustin,
``Duality of the
fermionic 2d black hole and N = 2 Liouville theory as  mirror symmetry,''
JHEP {\bf 0108}, 045 (2001)
[arXiv:hep-th/0104202]
%%CITATION = HEP-TH 0104202;%%
D.~Tong,
``Mirror mirror on the wall: On two-dimensional
black holes and Liouville  theory,''
JHEP {\bf 0304}, 031 (2003)
[arXiv:hep-th/0303151].
%%CITATION = HEP-TH 0303151;%%
}

%\JanikHB
\lref\JanikHB{
R.~A.~Janik,
``Exceptional boundary states at c = 1,''
Nucl.\ Phys.\ B {\bf 618}, 675 (2001)
[arXiv:hep-th/0109021].
%%CITATION = HEP-TH 0109021;%%
}

\lref\ShenkerUF{
S.~H.~Shenker,
``The Strength Of Nonperturbative Effects In String Theory,''
RU-90-47
%%\href{http://www.slac.stanford.edu/spires/find/hep/www?r=ru-90-47}{SPIRES entry}
{\it Presented at the Cargese Workshop on Random Surfaces, Quantum Gravity and Strings, Cargese, France, May 28 - Jun 1, 1990}
}

\lref\CardyIR{
J.~L.~Cardy,
``Boundary Conditions, Fusion Rules And The Verlinde Formula,''
Nucl.\ Phys.\ B {\bf 324}, 581 (1989).
%%CITATION = NUPHA,B324,581;%%
}

\lref\GreensiteYC{
J.~Greensite and M.~B.~Halpern,
``Stabilizing Bottomless Action Theories,''
Nucl.\ Phys.\ B {\bf 242}, 167 (1984).
%%CITATION = NUPHA,B242,167;%%
}

\lref\KarczmarekXM{
J.~L.~Karczmarek, H.~Liu, J.~Maldacena and A.~Strominger,
``UV finite brane decay,''
arXiv:hep-th/0306132.
%%CITATION = HEP-TH 0306132;%%
}

\lref\DouglasUP{
M.~R.~Douglas, I.~R.~Klebanov, D.~Kutasov, J.~Maldacena, E.~Martinec and N.~Seiberg,
``A New Hat For The c=1 Matrix Model,''
arXiv:hep-th/0307195.
%%CITATION = HEP-TH 0307195;%%
}

%\MartinecKA
\lref\MartinecKA{
E.~J.~Martinec,
``The annular report on non-critical string theory,''
arXiv:hep-th/0305148.
%%CITATION = HEP-TH 0305148;%%
}

\lref\AlexandrovNN{
S.~Y.~Alexandrov, V.~A.~Kazakov and D.~Kutasov,
``Non-perturbative effects in matrix models and D-branes,''
arXiv:hep-th/0306177.
%%CITATION = HEP-TH 0306177;%%
}

\lref\GaiottoYF{
D.~Gaiotto, N.~Itzhaki and L.~Rastelli,
``On the BCFT Description of Holes in the c=1 Matrix Model,''
arXiv:hep-th/0307221.
%%CITATION = HEP-TH 0307221;%%
}

%\HoravaJY
\lref\HoravaJY{
E.~Witten,
``D-branes and K-theory,''
JHEP {\bf 9812}, 019 (1998)
[arXiv:hep-th/9810188];
%%CITATION = HEP-TH 9810188;%%
P.~Horava,
``Type IIA D-branes, K-theory, and matrix theory,''
Adv.\ Theor.\ Math.\ Phys.\  {\bf 2}, 1373 (1999)
[arXiv:hep-th/9812135].
%%CITATION = HEP-TH 9812135;%%
}

%\SenMG
\lref\SenMG{
A.~Sen,
``Non-BPS states and branes in string theory,''
arXiv:hep-th/9904207.
%%CITATION = HEP-TH 9904207;%%
}

\lref\Kraus{
J.~A.~Harvey, P.~Horava and P.~Kraus,
``D-sphalerons and the topology of string configuration space,''
JHEP {\bf 0003}, 021 (2000)
[arXiv:hep-th/0001143].
%%CITATION = HEP-TH 0001143;%%
}

\lref\MikovicQF{
A.~Mikovic and W.~Siegel,
``Random Superstrings,''
Phys.\ Lett.\ B {\bf 240}, 363 (1990).
%%CITATION = PHLTA,B240,363;%%
}

%\OoguriCK
\lref\OoguriCK{
H.~Ooguri, Y.~Oz and Z.~Yin,
``D-branes on Calabi-Yau spaces and their mirrors,''
Nucl.\ Phys.\ B {\bf 477}, 407 (1996)
[arXiv:hep-th/9606112].
%%CITATION = HEP-TH 9606112;%%
}

\lref\arod{ J.~P.~Rodrigues and A.~J.~van Tonder,
%``Marinari-Parisi and supersymmetric collective field theory,''
Int.\ J.\ Mod.\ Phys.\ A {\bf 8}, 2517 (1993)
[arXiv:hep-th/9204061]; R.~de Mello Koch and J.~P.~Rodrigues,
%``The Collective field theory of a singular supersymmetric matrix model,''
Phys.\ Rev.\ D {\bf 51}, 5847 (1995) [arXiv:hep-th/9410012].}

\lref\freedman {D.~Z.~Freedman and P.~F.~Mende,
%``An Exactly Solvable N Particle System In Supersymmetric Quantum Mechanics,''
Nucl.\ Phys.\ B {\bf 344}, 317 (1990); L.~Brink, T.~H.~Hansson,
S.~Konstein and M.~A.~Vasiliev,
 %``The Calogero model: Anyonic representation, fermionic extension and
%supersymmetry,''
Nucl.\ Phys.\ B {\bf 401}, 591 (1993) [arXiv:hep-th/9302023].}

\lref\poly{ A.~P.~Polychronakos,
``Generalized statistics in one dimension,''
arXiv:hep-th/9902157. }
%%CITATION = HEP-TH 9902157;}

\lref\cmtricks{ P.~Desrosiers, L.~Lapointe and P.~Mathieu,
 %``Supersymmetric Calogero-Moser-Sutherland models: superintegrability
%structure and eigenfunctions,''
arXiv:hep-th/0210190;
%%CITATION = HEP-TH 0210190;%%
%\cite{Desrosiers:2003me}
%``Generalized Hermite polynomials in superspace as eigenfunctions of the
%supersymmetric rational CMS model,''
Nucl.\ Phys.\ B {\bf 674}, 615 (2003) [arXiv:hep-th/0305038];
%%CITATION = HEP-TH 0305038;%%
%\cite{Desrosiers:2002ww}
%``Jack polynomials in superspace,''
Commun.\ Math.\ Phys.\  {\bf 242}, 331 (2003)
[arXiv:hep-th/0209074].}
%%CITATION = HEP-TH 0209074;%%

\lref\BKZ{
E.~Brezin, V.~A.~Kazakov and A.~B.~Zamolodchikov,
``Scaling Violation In A Field Theory Of Closed Strings In One Physical Dimension,''
Nucl.\ Phys.\ B {\bf 338}, 673 (1990).
%%CITATION = NUPHA,B338,673;%%
}

\lref\GinspargIS{
P.~Ginsparg
and G.~W.~Moore, ``Lectures On 2-D Gravity And 2-D String
Theory,'' arXiv:hep-th/9304011.
%%CITATION = HEP-TH 9304011;%%
}

\lref\igorreview{
I.~R.~Klebanov, ``String theory in two-dimensions,''
arXiv:hep-th/9108019.
%%CITATION = HEP-TH 9108019;%%
}

\lref\kutasovniarchos{
A.~Sen, ``Dirac-Born-Infeld Action on the Tachyon
Kink and Vortex,'' [arXiv:hep-th/0303057];
%%%CITATION = HEP-TH 0303057;%%
%D.~Kutasov and V.~Niarchos,
%``Tachyon effective actions in open string theory,''
%arXiv:hep-th/0304045;
%%CITATION = HEP-TH 0304045;%%}
K.~Okuyama,
``Wess-Zumino term in tachyon effective action,''
JHEP {\bf 0305}, 005 (2003)
[arXiv:hep-th/0304108].
%%CITATION = HEP-TH 0304108;%%
}

%\BanksCY
\lref\BanksCY{
T.~Banks, L.~J.~Dixon, D.~Friedan and E.~J.~Martinec,
``Phenomenology And Conformal Field Theory Or Can String Theory Predict The Weak Mixing Angle?,''
Nucl.\ Phys.\ B {\bf 299}, 613 (1988).
%%CITATION = NUPHA,B299,613;%%
}

\lref\bergman{
O.~Bergman and M.~R.~Gaberdiel,
``Stable non-BPS D-particles,''
Phys.\ Lett.\ B {\bf 441}, 133 (1998)
[arXiv:hep-th/9806155].
%%CITATION = HEP-TH 9806155;%%
}

\lref\DJ{
S.~R.~Das and A.~Jevicki,
``String Field Theory And Physical Interpretation Of D = 1 Strings,''
Mod.\ Phys.\ Lett.\ A {\bf 5}, 1639 (1990).
%%CITATION = MPLAE,A5,1639;%%
}

\lref\sena{A.~Sen, ``Stable non-BPS bound states of BPS D-branes,''
\jhep{9808}{98}{010}, [arXiv:hep-th/9805019];
``SO(32) spinors of type I and other solitons on brane-antibrane pair,''
\jhep{9809}{98}{023}, [arXiv:hep-th/9808141];
``Type I D-particle and its interactions,''
\jhep{9810}{98}{021}, [arXiv:hep-th/9809111].
%``Non-BPS states and branes in string theory,''
%[arXiv:hep-th/9904207],
%``BPS D-branes on non-supersymmetric cycles,''
%\jhep{9812}{98}{021}, hep-th/9812031,
}

%\BilloTV
\lref\Billo{
M.~Billo, B.~Craps and F.~Roose,
``Ramond-Ramond couplings of non-BPS D-branes,''
JHEP {\bf 9906}, 033 (1999)
[arXiv:hep-th/9905157].
%%CITATION = HEP-TH 9905157;%%
}

\lref\Kennedy{
C.~Kennedy and A.~Wilkins,
``Ramond-Ramond couplings on brane-antibrane systems,''
Phys.\ Lett.\ B {\bf 464}, 206 (1999)
[arXiv:hep-th/9905195].
%%CITATION = HEP-TH 9905195;%%
}

\lref\SenBdy{
A.~Sen, ``Stable non-BPS bound states of BPS D-branes,''
\jhep{9808}{98}{010}, hep-th/9805019.}

\lref\BDLR{
D.~Gepner,
``Space-Time Supersymmetry In Compactified String Theory And Superconformal
Models,''
Nucl.\ Phys.\ B {\bf 296}, 757 (1988);
%%CITATION = NUPHA,B296,757;%%
A.~Recknagel and V.~Schomerus,
``D-branes in Gepner models,''
Nucl.\ Phys.\ B {\bf 531}, 185 (1998)
[arXiv:hep-th/9712186];
%%CITATION = HEP-TH 9712186;%%
I.~Brunner, M.~R.~Douglas, A.~E.~Lawrence and C.~Romelsberger,
``D-branes on the quintic,''
JHEP {\bf 0008}, 015 (2000)
[arXiv:hep-th/9906200].
%%CITATION = HEP-TH 9906200;%%
%\lref\Nepomechie{
R.~I.~Nepomechie,
``Consistent superconformal boundary states,''
J.\ Phys.\ A {\bf 34}, 6509 (2001) [arXiv:hep-th/0102010].
%%CITATION = HEP-TH 0102010;%%
%\GaberdielZQ
%\lref\GaberdielZQ{
M.~R.~Gaberdiel and A.~Recknagel,
``Conformal boundary states for free bosons and fermions,''
JHEP {\bf 0111}, 016 (2001)
[arXiv:hep-th/0108238].
%%CITATION = HEP-TH 0108238;%%
}

\lref\AganagicDB{
M.~Aganagic, A.~Klemm, M.~Marino and C.~Vafa,
``The topological vertex,''
arXiv:hep-th/0305132.
%%CITATION = HEP-TH 0305132;%%
}

\lref\stherm{
A.~Maloney, A.~Strominger and X.~Yin, ``S-brane thermodynamics,''
arXiv:hep-th/0302146.
%%CITATION = HEP-TH 0302146;%%
}

\lref\IvanovWP{
E.~A.~Ivanov and S.~O.~Krivonos,
``U(1) Supersymmetric Extension Of The Liouville Equation,''
Lett.\ Math.\ Phys.\  {\bf 7}, 523 (1983)
[Erratum-ibid.\  {\bf 8}, 345 (1984)].
%%CITATION = LMPHD,7,523;%%
}

%\PolchinskiMT
\lref\PolchinskiMT{
J.~Polchinski,
``Dirichlet-Branes and Ramond-Ramond Charges,''
Phys.\ Rev.\ Lett.\  {\bf 75}, 4724 (1995)
[arXiv:hep-th/9510017].
%%CITATION = HEP-TH 9510017;%%
}

\lref\wittenmf{
E.~Witten,
``Constraints On Supersymmetry Breaking,''
Nucl.\ Phys.\ B {\bf 202}, 253 (1982).
%%CITATION = NUPHA,B202,253;%%
}

%\EguchiIK
\lref\Eguchi{
T.~Eguchi and Y.~Sugawara,
``Modular bootstrap for boundary N = 2 Liouville theory,''
JHEP {\bf 0401}, 025 (2004)
[arXiv:hep-th/0311141].
%%CITATION = HEP-TH 0311141;%%
}

%\AhnTT
\lref\Ahn{
C.~Ahn, M.~Stanishkov and M.~Yamamoto,
``One-point functions of N = 2 super-Liouville theory with boundary,''
arXiv:hep-th/0311169.
%%CITATION = HEP-TH 0311169;%%
}

\Title{\vbox{\baselineskip12pt
\hbox{hep-th/0308105}\hbox{PUPT-2095} }} {\vbox{ {\centerline
{Two-dimensional Superstrings and the}}\vskip .2in {\centerline
{Supersymmetric Matrix Model}} }}

\centerline{John McGreevy, Sameer Murthy
and Herman Verlinde}
\smallskip
\centerline{\sl Department of Physics,  Princeton University}
\centerline{\sl Princeton, NJ 08544, USA}
\medskip

\vskip .3in
\centerline{\bf Abstract}

We present evidence that the supersymmetric matrix model of
Marinari and Parisi represents
the
world-line theory of $N$ unstable D-particles in
type II superstring theory in two dimensions.
This identification suggests that the matrix
model gives a holographic description of superstrings in a
two-dimensional
black hole geometry.

\Date{August, 2003}

\newsec{Introduction}

Matrix models provide an elegant and powerful formalism for
describing low-dimensional
%non-critical
string theories. Recently,
it was proposed that the large $N$ matrix variables
can be viewed
as the modes of
$N$ unstable D-particles in
the corresponding string theory,
in a decoupling limit
\refs{\McGreevyKB,\MartinecKA}. This proposal reinterprets the
matrix-model/string-theory
correspondence
as a holographic open-string/closed-string duality,
and suggests a search algorithm for
more examples.
It has been clarified
\refs{\KlebanovKM,\McGreevyEP,\AlexandrovNN,\GaiottoYF}
and very recently
extended to type 0 strings
\refs{\TakayanagiSM,\DouglasUP}.
In this note, we apply this perspective to
shed some new light on the physical identification of the
supersymmetric matrix model of Marinari and Parisi
\MarinariJC.

The paper is organized as follows. We begin by recollecting the
basic features of the Marinari-Parisi model and its
proposed continuum limit. In section 3, we review some of the
target space properties of 2-d superstring theory. In section 4 we
collect a list of correspondences between the two theories. Most
notably, we find that the open string spectrum on unstable
D-particles of the 2-d string theory is that of (a minor
improvement of) the MP model, expanded around the maximum of its
potential. We also make a direct comparison between the vacuum
structure and instantons of both models. We end with some
concluding remarks and open problems. Some technical discussions
are sequestered to appendices A and B.

\newsec{The Marinari-Parisi model}

The Marinari-Parisi model is the quantum mechanics of an $N\times
N$ hermitian matrix in a one-dimensional superspace \eqn\sfield{
{\Phi}(\tau,\theta,\bar\theta) = {M}(\tau) +
\bar\theta{\Psi}(\tau) + \bar{\Psi}(\tau)\theta + \theta\bar\theta
{ F}(\tau).} The action is \eqn\act{ S = -N \int
d\tau\,d\bar\theta\,d\theta\, {\rm Tr}\biggl\{ {1\over 2} \bar D
{\Phi} D {\Phi} + W_0({\Phi}) \biggr\}, } where $D, \bar D$ are
superspace derivatives.
%whose form is given in appendix A.
We can
choose a cubic superpotential \eqn\supo{ W_0({\Phi}) = {1\over 2}
{\Phi}^2 - {1\over 3} \lambda^2 {\Phi}^3.} The Feynman graph
expansion for the model generates a discretization of random
surfaces in superspace. Related work on supersymmetric matrix
models includes \refs{\GreensiteYC,\MikovicQF,\atish,\CohnZJ,\vanTonderVC,\RodriguesBY,\KarlinerCD,\GonzalezYU,\FerrettiFU,\BrusteinNC,\Feinberg}.

The model we will discuss is actually a slight modification of the
original MP model.
We will take the derivatives appearing in \act\
to be covariant with respect to gauge transformations
which are local in superspace;
their form is described in appendix A.  In the case of the $c=1$ matrix
model, its identification with the worldline theory of D-particles
made clear that the $U(N)$ conjugation symmetry of the matrix
model should be gauged. As we will see, the same correspondence in
our case suggests that we should introduce a superfield gauge
symmetry in the model \act, which naturally effects the
truncation to singlet states \atish.
%Details of this projection are given in appendix A.

The model with superpotential \supo\  has two classical
supersymmetric extrema $W_0'({\Phi})\!=\! 0$. These are minima of
the bosonic
potential ${V(M) = M^2(1 - \lambda^2 M)^2}$. In addition,
$V$ has an unstable critical point at $M_c = {1\over 2\lambda^2}$.
The quadratic form of the action, when expanded near this
non-supersymmetric critical point is (defining $Y= M-M_c$)
\eqn\quapot{S = -N \int d\tau {\rm Tr}\biggl\{ {1\over 2} (D_\tau
Y)^2 + \bar\Psi D_\tau \Psi  + {1\over 2} \lambda^2 Y^2 -{1\over
16\lambda^2} \biggr\}.}
In the following we will argue that this
action can be viewed as that of $N$ unstable D-particles,
localized in the strong coupling/curvature region of the 2d
string theory background.

In the MP model, the fermi level is not an independent parameter,
in that it is determined by the form of the potential \atish.
%The MP model does not have an
%adjustable fermi level \atis
Critical behavior arises instead in this model through a
singularity in the norm of the ground-state wavefunction \atish.
%in eigenvalue space \atish.
We will discuss the ground states of the matrix model further in
\S4, but for now it suffices to study an exemplary one,
$\ket{f_0}$, whose norm is given by \eqn\divergingnorm{ e^{ \CF }
\equiv | f_0 |^2 =
%\int \prod_i dz_i ~e^{ - 2 W_{\rm eff}(z)},
\int \prod_i dz_i ~\prod_{i < j} (z_i - z_j)^2
~e^{ - 2 W_{0}(z)};
}
this is a $ c < 1$ matrix integral.
For odd $W_0$, there is an irrelevant divergence at
large $|z|$ which we simply cut off.
A critical limit
arises by tuning the potential $W_0$
to the $m=2$ pure-gravity critical point of \BIPZ,
near which the tree-level
free energy is $\CF \propto \kappa^{-2},$
with $\kappa^{-1} = (\lambda-\lambda_c)^{5/4} N $
providing the string coupling.

This limit naively gives a supersymmetric sigma model on
1d superspace $(\tau, \theta, \bar\theta)$ coupled to 2d
Liouville supergravity.
%However,
%the Marinari-Parisi model has not been as extensively studied
%as its bosonic counterpart, in
%part because no clear candidate 2d superstring theory could be
%identified. Furthermore,
In \chaudpol, however, the following argument was presented
against such a
description of the continuum limit: the matter part of the
action is necessarily interacting, and has a one-loop beta
function predicting that the coupling grows in the IR and that the
matter fields become disordered. This would seem to indicate that
the superspace coordinates $(\tau, \theta, \bar\theta)$ become
massive, and that spacetime
%supersymmetry
does not survive in the
critical theory. We consider this conclusion premature. The
reasoning assumes that the matter theory and the worldsheet
gravity are coupled only via the gauge constraints. This is not
the case for the supersymmetric string in two dimensions \KS.

\newsec{Two-dimensional Superstrings}

To formulate 2d superstring theory, one starts from $\CN=2$
Liouville theory \IvanovWP\ and then performs a consistent GSO
projection to obtain a string theory with target space
supersymmetry \refs{\KS,\sameer}. Unlike bosonic and $\CN=1$
supersymmetric Liouville theory, the time direction $\tau$ is
involved in the $\CN=2$ supersymmetry algebra, and it is involved
in the $\CN=2$ Liouville interaction as well:
 \eqn\sinliouv{\CL^{SL}_{int} = \psi \tt \psi
\, e^{-{1 \over 2}( \rho +\tt{\rho} + i( \tau -\tt \tau))} + c.c.}
with $\psi = \psi_\rho + i \psi_\tau$.

Interestingly, $\CN=2$ Liouville theory has been shown
\HoriAX\ to be dual
to superstrings propagating inside the 2d black hole
defined by the supercoset $SL(2,\IR)/U(1)$ \wittenbh.
The semiclassical background is
\eqn\metric{\eqalign{ & ds^2= d\rho^2+\tanh^2\rho d\tau^2, \quad
\qquad \tau \equiv \tau +{2\pi}; \cr & \Phi = \Phi_0 - \log \cosh
\rho,}} with $g_s=e^\Phi$.\foot{Type IIB string theory based on
this CFT is also equivalent, via a more trivial T-duality, to type
IIA string theory on the circle of the inverse radius, with
\sinliouv\ replaced by the corresponding momentum condensate.}
This explicitly shows that in the infrared region of small
$e^\rho$, the $\tau$ direction degenerates. The dependence of the
string background on $\rho$ can be attributed to a gravitational
dressing of the operators. This does however not preclude the
existence of a two-dimensional
continuum description (\cf\ the above discussion of
the scaling of the MP matrix model).

%start samedit

%\subsubsec{Symmetries and Spectrum}

We now summarize a few properties of the target space theory. A
good starting point is the worldsheet description of the Euclidean
theory, the fermionic cigar at the free-fermion radius \sameer.
The string worldsheet theory on the cigar has three conserved
currents: the left-moving and right-moving chiral currents $J$ and
$\tt J$, with
 \eqn\ntwowss{J=-\bar\psi \psi   +i2\partial \tau
\equiv i\partial ( H  + 2 \tau ) , } and a non-chiral current
whose integral charge $P_\tau$ is the quantized euclidean energy,
\ie\ the discrete momentum around the cigar. The chiral projection
that defines the type II theories is the condition that physical
operators should have a local OPE with the spectral flow
operators, which in type IIB string theory  takes the form
 \eqn\defs{S= e^{-{1\over 2}{\varphi} + {i \over 2}
(H+ 2\tau)}
\qquad \qquad \tt S=
 e^{-{1\over 2}{\tt\varphi} + { i\over 2}(\tt H +2 \tt  \tau)}}
(here $\varphi$ denotes the bosonized superghost current).
Since the $U(1)$ $R$ current of the ${\cal N}=2$
algebra involves the compact boson $\tau$ in addition to the
worldsheet fermions, the symmetry generator $P_\tau$ is an $R$ symmetry
(here $\CS =  \int dz ~S$ is the supercharge)
 \eqn\pssbaralg{
  [P_\tau, \CS]= \half \CS\qquad \qquad   [P_\tau, \tt \CS]= \half \tt \CS\, .}
States in a given supersymmetry multiplet therefore do not all
have the same energy.
%The conjugate charges $\bar \CS, \bar{\tilde \CS}$,
%which carry negative $\theta$-momentum,
%are not visible on the worldsheet,
%but are surely present in a
%second quantized description of the theory.

The perturbative closed string spectrum in the euclidean IIB
string theory consists \sameer\
of an NSNS (non-tachyonic) tachyon with odd
winding modes, a left-moving periodic RR scalar (the self-dual
axion), and a right moving complex fermion $\Upsilon$ with
half-integer momenta.  This is the expected behaviour for spinors
which are single valued on the cigar. Therefore, in the compact
theory, a rotation $\tau \to \tau + 2 \pi $ acts on the spacetime
fields as
%\eqn\rotation{
$ e^{ 2 \pi i P_\tau}  = (-1)^{F_s}$,
%}
where $F_s$ is the target-space fermion number. There is another
$\IZ_2$ symmetry $(-1)^{F_L}$ which acts by \eqn\mflaction{ \chi
\mapsto - \chi, ~~~ \Upsilon\mapsto \bar \Upsilon. }

The $\CN=2$ Liouville which has primarily been considered in the
literature has a euclidean time direction. On the other hand,
matrix quantum mechanics is most easily described in a real-time
Hamiltonian language. It will therefore be convenient for us to
hypothesize a consistent analytic continuation of this theory.
%of the $\tau$ direction
%of the Liouville theory, with a sinh-Liouville type of
%interaction.
However, this analytic continuation needs to be understood better.
Translating
to a Minkowskian spectrum, we find a left-moving scalar $\chi$ and
a complex right-moving fermion $\Upsilon$.  In addition to these
propagating degrees of freedom, there are also discrete physical
states at special energies.

%This can be seen as follows.
%The Euclidean blackh hole metric is
%of the form SOMETHING LIKE \eqn\euclindeanbhmetric{
%ds^2  = f( uv ) du dv , ~~~f(r) = {1 \over 1 + r^2}
%}
%where $u,v$ are holomorphic coordinates related to $\rho,
%\tau$ by\eqn\holccoords{
%u = e^{ \rho + i \tau} , ~~v =  e^{ \rho - i \tau} .
%}
%OR
%\eqn\lightconecoords{
%x^\pm = e^{ \rho \pm \tau} ~~?
%}
%This can be analytically continued
%to a Minkowski black hole as
%\eqn\bhmetric{
%ds^2  = f( x^+x^- ) dx^+ dx^- , ~~~f(r) = {1 \over 1 - r^2}
%}
%Define
%\eqn\timeandspace{
%\sigma \equiv {1 \over \sqrt 2} ( x^+ + x^- ) = \sqrt 2
%e^{ - \rho} \sinh \tau , ~~~
%t \equiv {1 \over \sqrt 2} ( x^+ - x^- ) = \sqrt 2
%e^{ - \rho} \cosh \tau .}
%A collection of spacelike hypersurfaces is then
%defined by taking constant $ t$ slices.
%The quantity \eqn\scharge{ Q_s= \int_{ \Sigma(t)} \star d \phi
%= \int_{ \Sigma(t)} \star \left( \del_+ \chi dx^+
%+ \del_- U dx ^- \right)
%Q_s = \int_{  \Sigma(t) }\left( \del_+ \chi dx^+ - \del_- U dx^-
%\right)
%\left( \chi - U \right)_{\del \Sigma(t) }
%= e^{ - \rho} \left(  (p_L - p_R) \cosh \tau
%+ ( p_L + p_R ) \sinh \tau \right) \big|^{ \rho = 0 } _
%{\rho = \infty}
%= p_L +  p_R  } is conserved in the absence of sources for $\phi$.
%\foot{While $ \int d \phi $ measures the net sD-brane charge, $
%\int \star d \phi$ measures sD PLUS anti-sD-brane charge. Note
%that there is a special value ($p = 0$) where there are equal
%numbers of sD-branes and anti-sD-branes. This particular state is
%mapped to itself by the action of $(-1)^{F_L}$. }

\subsubsec{D-Instantons and Flux sectors}

The physics of the RR axion is closely linked to that of
D-instantons.
%, see e.g. \Kraus.
Two-dimensional IIB string theory, however, has some special
features. First, the axion is a self-dual middle-rank form; it
couples both electrically and magnetically to the D-instanton. One
important implication of this is that the axion itself does not
have a well-defined constant zeromode. Secondly, unlike the 10d
case, where the BPS D-instanton breaks sixteen supercharges and
thus carries an even number of fermion zeromodes, it seems that
the 2d D-instanton only breaks {\it one} supersymmetry and
therefore carries only one fermion zeromode. It thus interpolates
between sectors with opposite fermion parity. A preliminary study
of the D-instanton boundary state in appendix B bears this out.

%[FACTORS OF $2\pi$??]
%Thinking of the axion as a chiral scalar,
%the operator which inserts n D-instantons
%is the chiral vertex operator
%$ e ^{ i n \chi (x_+)}$.
%A preliminary study of the
%D-instanton boundary state
%indicates that it has only one fermion
%zeromode.  This

At this point it is natural to introduce the right-moving scalar
field $U$ via bosonization  $\Upsilon = e^{ i U}$. In this
bosonized language, the entire field-theoretic spectrum of 2d type
IIB can thus be reassembled into a single non-chiral scalar field
$\phi = \phi_L + \phi_R$ with \eqn\phidef{ \phi_L = \chi,\qquad
\phi_R = U.} Since $\phi_R=U$ is periodic with the free-fermion
radius, it is natural to suspect that the axion $\phi_L=\chi$ is
periodic with the free-fermion radius as well.\foot{This claim can
be verified (or refuted) by computing the axion charge carried by
a D-instanton \PolchinskiMT, or the axion flux produced by a
decaying D-brane, along the lines of
\refs{\KlebanovKM,\DouglasUP}. }  Note that $\mfl$ as defined
above acts by $ \mfl: \phi\mapsto - \phi$.

The fact that the D-instanton has only one fermion zero mode means
the operator that creates it carries fermion number 1. This
indicates that $ \oint_ \gamma \bar \del U = 1 $, where $\gamma$
is a contour containing the instanton. Further, from the coupling
of the D-instanton to the RR axion, we expect that in the presence
of a D-instanton
%\eqn\fluxchange{
$\oint_\gamma \del \chi$ = 1. A natural candidate for the
effective operator with the right properties to create a
D-instanton at the space-time location $x$ is then
\eqn\instantoncreator{
%:e^{ i \phi (x_+,x_-) }:
%= : e^{ i \chi(x_+)}:~: e^{ i U (x_-)} :.
e^{ i \phi (x) } = e^{ i(\chi(x)+ U (x))} . }

 Instantons are
tunnelling events that interpolate between perturbative sectors.
These sectors are characterized by an integer flux (here $\Sigma$
denotes a space-filling contour): \eqn\charge{\int_\Sigma
\partial_0 \phi = k  \qquad \quad k \in {\bf Z},}
which is the quantized momentum dual to the constant zero mode
$\phi_0$ of $\phi$ ($\phi_0$ is periodic with period $2\pi$). The
integer $k$ can be thought of as a slight generalization of the
``s-charge'' of \stherm.\foot{ Since $\del_+ \phi = \del_+ \chi$,
sectors with nonzero $p_L$ are backgrounds in which flux quanta of
the RR axion are turned on. Backgrounds of two-dimensional type
IIA strings with RR flux were described in \BerkovitsTG.}
%By this we mean
%a lightlike gradient
%\eqn\efield{
%\chi (\tau, \rho) = E x_+
%}
%where asymptotically $x _+ = \rho + \tau$.
%[THE DEFINITION OF $x_+$ IS OPEN TO DEBATE.]
%Thinking of the axion as a chiral scalar in two
%dimensions,
%the vacuum with $E$ units of flux is
%simply the one with
%left-moving momentum $p_L = E$.
%Note that because $\chi$ is self-dual,
%the RR flux $E$ appearing in \efield, and the axion zeromode
%$\chi_0 \equiv \int dx^- {\del \over \del x^-} \chi $
%are canonically conjugate variables.

\newsec{Dual Correspondence}

We will  now try match the physics of the Marinari-Parisi matrix
model with that of the two-dimensional type IIB string theory.
Following the logic of \refs{\McGreevyKB,\KlebanovKM} we start
by examining the open string spectrum of the unstable
D-particles.

\subsubsec{D-particles}

In type II string theory, the boundary state for an unstable
Dp-brane has the form \refs{\bergman,\sena,\SenMG,\GaberdielJR}
\eqn\unstable{
%\sqrt 2
\ket{ \hat{Dp}} = \ket{B, NSNS; +} - \ket{ B, NSNS; -}. } Here
$\ket{B, NSNS; \eta} $ denotes a boundary state in the NSNS
sector, satisfying
%$(G^\pm_r- i \eta \tilde G ^\pm_r) \kket { B, \eta} = 0 $.
$(G_r- i \eta \tilde G_r) \ket { B;\eta} = 0$; $G = G^+ + G^-$ is
the gauged worldsheet supercurrent. The boundary state describing
an unstable brane with unperturbed tachyon
contains no term built on Ramond primaries.
Experience with less supersymmetric Liouville models suggests that
branes localized in the Liouville direction correspond to boundary
states associated with the Liouville vacuum state, which we will
call $\ket{B_0;\eta}$. The defining property of these states is
that the corresponding bosonic open string spectrum (of NS-sector
open strings for which both end-points satisfy this specific
boundary condition) have support only at Liouville momentum $P = -
i $, corresponding to the identity Liouville state.

Details regarding these boundary states appear in appendix B.
Using the general formula \unstable, the basic unstable D0-brane
of type IIB is represented by the boundary state
%\eqn\unstabledzero{
%\sqrt 2 ~
$ \ket{ \hat{D0}} = \ket{ B_0; + } - \ket {B_0; -}. $ Study of the
annulus amplitude for this D-brane, detailed in appendix B,
reveals that the open string spectrum on this brane is precisely
that of the Marinari-Parisi model, expanded as in \quapot,
including the gauge supermultiplet. This correspondence is the
first strong indication that the MP model describes the type IIB
non-critical string theory.

\subsubsec{Symmetry considerations}

%\subsubsec{Matching of symmetries}
%AT THIS POINT YOU'VE ALREADY DISCUSSED $F$.
There are two prominent continuous symmetries of the MP model:
there is the conserved energy $H$, and there is the overall
fermion number
%\eqn\fermionumber{
$\hat F \equiv \sum_i \psi_i^\dagger \psi_i.$
%}
In the Hilbert space of the MP model, the quantum number $F$ takes
$N$ different values, for which we take the CP-invariant choice $
-N/2, \dots, N/2$. We would like to identify $(-1)^{F_s}$ of the
target space theory with $ (-1)^F$
%$F \equiv \sum_i \psi_i^\dagger \psi_i$
of the MP model. Further, as in the bosonic and type 0 cases, we
identify the Hamiltonians of the systems
$$ H = P_\tau .$$

The matrix model can also have a $\IZ_2$ R-symmetry. Its
interpretation can be understood as follows. Due to the coupling
between the D-brane worldvolume fields and the closed strings, the
worldvolume fields transform under $(-1)^{F_L}$. $\mfl$ acts by
\refs{\sena,\Kraus} \eqn\mflactiononwv{ Y \leftrightarrow - Y,~~~
\psi \leftrightarrow \psi^\dagger.} We will see below that this is
consistent with $\Upsilon \leftrightarrow \Upsilon^\dagger$.
Therefore $\mfl$ acts as an R-symmetry in the matrix quantum
mechanics: it acts on the superspace coordinates as $\mfl: \theta
\leftrightarrow \bar \theta.$ In order for this to be a symmetry
of the worldline action, $W_0$ {\it must be an odd function of}
$Y$. This implies that supersymmetry is broken, since then none of
the standard \wittenmf\ candidate supersymmetric ground states
$e^{ \pm W} \ket{0} $ is normalizible.

\subsubsec{Spectrum and $c=1$ Scaling}

%\eqn\antisym{ f_0( \bfz) = (-1)^{\sigma} f_0(\sigma(\bfz)) }
%where
%$\sigma(z)$ denotes a permutation of the eigenvalues, and
%$(-1)^\sigma$ is the determinant of the permutation.
In \atish, it was shown that the supercharges act within the
space of super matrix eigenvalues as \eqn\hermit{ Q = \sum_k
\psi_k^\dagger\Bigl({1\over N} {\partial\over{\partial
 z_k}} +
       {{\partial W_{\rm eff}( z)}\over{\partial z_k}}\Bigr)\, , \qquad \quad
 Q^\dagger = \sum_k \psi_k\Bigl({1\over N} {\partial\over{\partial
 z_k}}
             - \,{{\partial
             W_{\rm eff}( z)}\over{\partial z_k}}\Bigr),}
where \eqn\weff{ W_{\rm eff}(z) = \sum_k W_0(z_k) - {1\over N}
\sum_{k<l} \log (z_k-z_l). } These supercharges exactly coincide
\arod\ with those of the supersymmetric Calogero-Moser model
\freedman\cmtricks. The corresponding Hamiltonian reads
\eqn\cmham{ {\CH} = \sum_{i=1}^N \left({1\over 2 } p_i^2+ V(z_i) +
{2\over N} W_0^{''}(z_i) \psi_i^\dagger \psi_i \right) + {1\over
N^2} \sum_{i<j} {1- \kappa_{ij}\over (z_{i}-z_j)^2}}
%-\frac{N}{2}[\beta(N-1)-\omega]
where with  $p_i = {-i\over N} {\partial \over \partial z_i}$ and
$\kappa_{ij} = 1-(\psi_i-\psi_j)(\psi^\dagger_i-\psi^\dagger_j)$
is the fermionic exchange operator \cmtricks; it assigns
fermi-statistics to the spin-down eigenvalues, and bose-statistics
to the spin-up ones (here we are using the terminology of appendix
A). The potential reads \atish \eqn\pot{V(z) = {1\over
2}\Bigr(W_0^{'}(z)\Bigl)^2 - \, W_0^{''}\!(z),} where we used that
for a cubic superpotential $W_0$ one has $
\sum_{i<j} {W_0'(z_i)- W_0'(z_j) \over z_i -z_j} = {(N-1)}%\over N}
\sum_i W_0^{''}\!(z_i)$. We see that the Hamiltonian describes a
system of interacting eigenvalues. The interaction is such that
eigenvalues always repel each other: it represents a $2/r^2$
repulsion between the boson states with $\kappa_{ij}=-1$, and
although it vanishes between two fermionic states with
$\kappa_{ij}=1$, such particles still avoid each other
since their wavefunctions are antisymmetric.

In the case all the eigenvalues have spin down, so that all
$\kappa_{ij}=1$, the Hamiltonian reduces to a decoupled set of
one-particle Hamiltonians. It is easy to write the ground state
wave function in this case. Let us introduce the notation
$\Delta(i_1,..i_k) = \prod_{i < j \in (i_1,..i_k)} (z_{ij})$, the
vandermonde of the $k$ variables $\{z_{i_k}\}$ (here $z_{ij} =
z_i-z_j$). In this notation: \eqn\fnot {|f_0 \rangle \equiv f_0
\downvac = e^{\rm{Tr}W_0} \Delta(1,2..n) \downvac.}
%f_0(z) =%\prod_{1\leq j<k\leq N}\left( z_{jk}\right)
%\Delta(z)^\beta e^{-{N\over 2} \omega_0\sum_{i=1}^N z^2_i} \, .
This vacuum state represents the filled Fermi sea of the first $N$
energy levels. In the harmonic potential, this is a supersymmetric
ground state. In the cubic potential, there is a corresponding
ground state in each well, only one of which is perturbatively
supersymmetric.

In the sectors with non-zero fermion number, there are no
normalizable supersymmetric ground states, even for the harmonic
well. In this case, one can show that the groundstate eigenfunction in the
fermion number $k$ sector is: \eqn\ngndst {| f_k \rangle =
\sum_{i_1< i_2 <..< i_k} \Delta(i_1,..i_k) \prod_{m=1}^k
\psi^\dagger_{i_m} | f_0 \rangle.} This wavefunction obeys the
right statistics imposed by gauge invariance. From the expression
\ngndst\ we can read off that the energy levels of the spin up
eigenvalues are double spaced relative to that between spin down
states \poly.
\doublefig\potentialdiagram{
The approximate probe superpotential, superpotential 
and bosonic potential of the $f_0$ state.
The fermi level of the
perturbatively supersymmetric ground state $f_0$ coincides with
the minimum of the right well ({\it left}). The ground state in the sector
with $k$ spin up eigenvalues fills up to the $N+k$-th energy
level, which in the double scaling limit approaches the unstable
maximum ({\it right}).
}{\epsfxsize2.3in\epsfbox{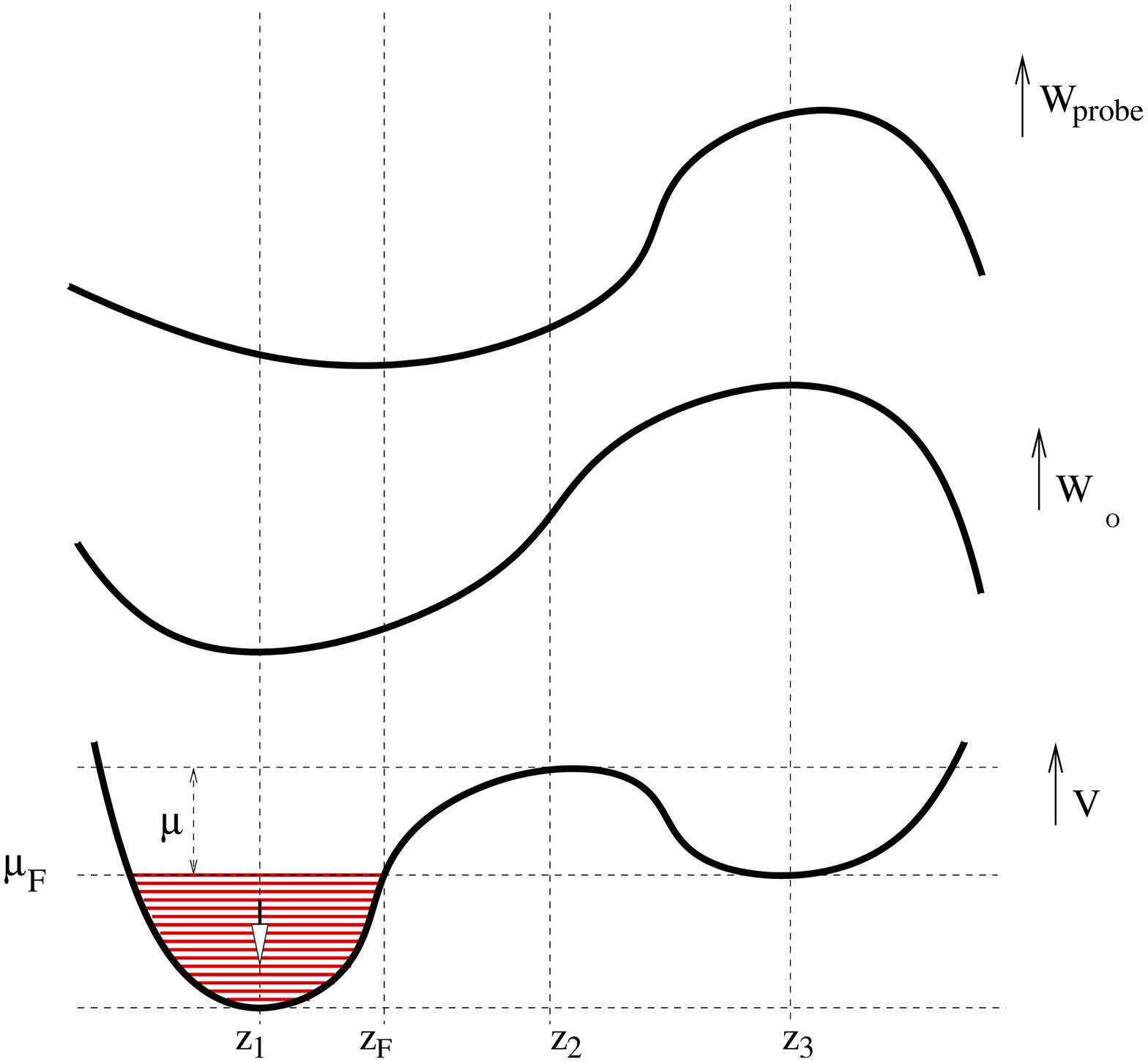}}
{\epsfxsize2.0in\epsfbox{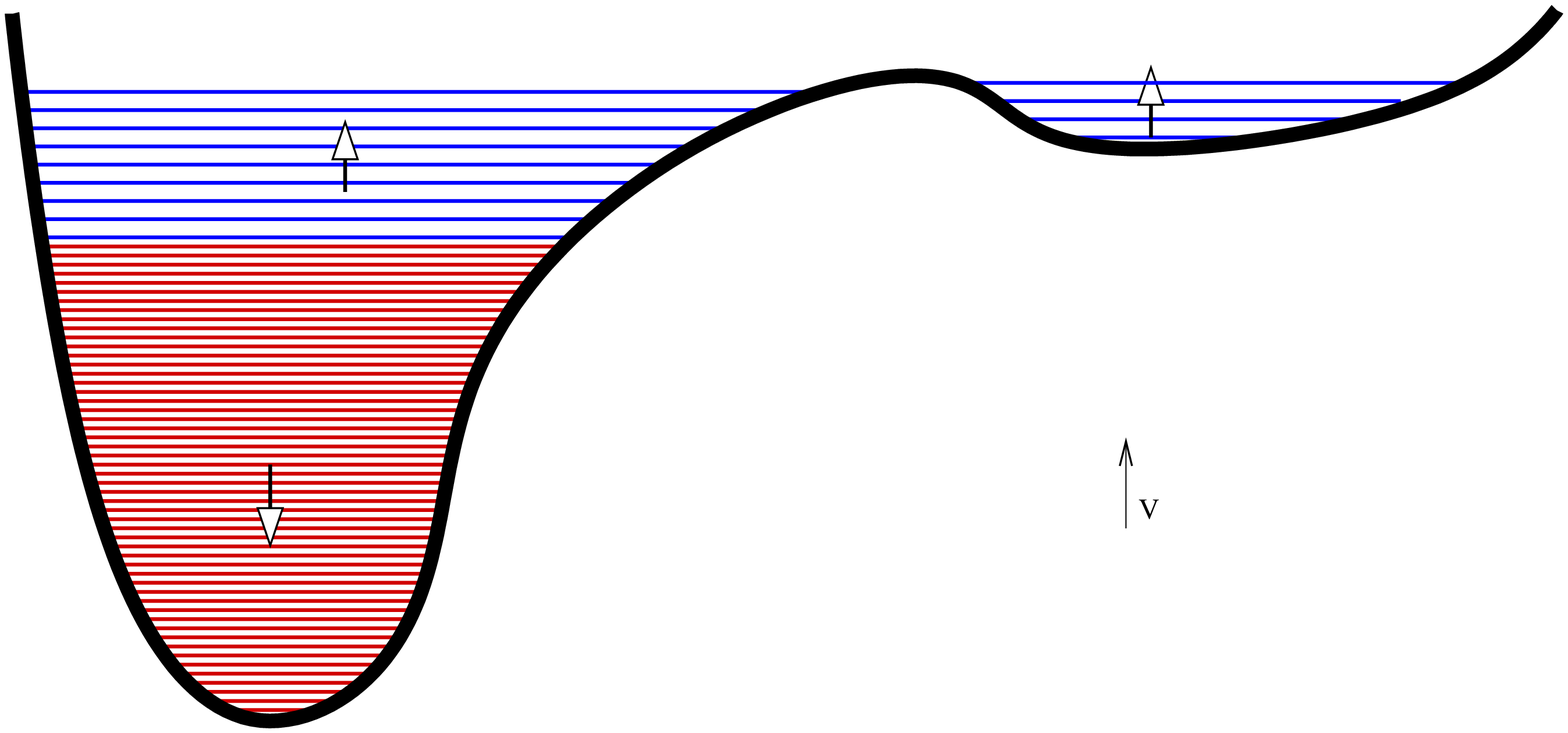}}
Now let us discuss the system with a cubic superpotential. The
state $f_0$ is a perturbatively supersymmetric state, with a
filled fermi sea in the left well. It turns out that the fermi
level exactly coincides with the bottom of the second well, as
indicated on the left of Fig.\ 1. Although the super-Calogero-Moser model with
a cubic potential has no known exact treatment, it seems
reasonable to assume that the super-eigenvalues in each of the two
potential wells behave qualitatively similar as for the single
harmonic potential. This suggests that in the sector with $k$
up-spins, the ground state is well-approximated by taking the
single particle energy spectrum, and fill the first $N-k$ levels
with the fermionic eigenvalues, and all levels $N-k+2m$ with $m=1,
\ldots, k$ up to the $N+k$-th energy level. When $k$ gets large
enough, this $N+k$-th energy level starts approaching the unstable
maximum of the potential, as indicated on the right of Fig.\ 1. Here we expect
to find $c=1$ critical behavior. Our proposal is that the double
scaled eigenvalue dynamics near this unstable maximum encodes the
scattering non-perturbative physics of non-critical IIB strings.

\subsubsec{Matrix model instantons}

%Next, we will show in the matrix model that a

Single-eigenvalue tunneling events in the matrix model interpolate
between the sectors with different number of up-spins: they relate
the adjacent ground states $\ket{f_k}$ and $\ket{f_{k+1}}$. Note
that these sectors have opposite parity of fermion number.
%$\Delta
%F = 1$.
%This can be demonstrated by examining the quantum
%mechanics of an $(N+1)$st probe super-eigenvalue $Z$.
The probe super-eigenvalue $Z$ moves in the mean-field
superpotential \eqn\probesuperpotential{ W_{\rm probe}(Z) = W_0(Z)
-\half \sum_{i} \ln( Z - X_i ) . }
%For our purposes, we can approximate
%the repulsive
%effects of the sea of other eigenvalues
%by a quartic potential $V_{\rm probe}$
%with a minimum at $z_F$.
%This is produced by
%a cubic superpotential $W_{\rm probe}$
%which has
%a minimum at $z_F$.
%The two ground states of the probe are localized respectively in
%the two wells of the quartic potential $V_{\rm probe}$,
%qualitatively indicated in fig 1.
%Defining $ \psi_Z \ket{\uparrow} = 0$,
%and using
%\eqn\probesupercharges{
% Q_z = i\psi_Z (\del_z - \del_z W_{\rm probe}) , ~~
%Q_z^\dagger = i \psi_Z ( \del_z + \del_z W_{\rm probe}), }
%the two supersymmetric ground states of the probe
%are
%\eqn\probestates{
%\ket{+} \propto e^{- W_{\rm probe}(z)} \ket{\uparrow}, ~~
%\ket{-} \propto e^{+ W_{\rm probe}(z)} \ket{\downarrow}.
%}
%The key point now is simply that
%the wavefunction of $\ket{-}$
%is localized in the left well,
%while that of $\ket{+}$ is
%localized in the right well.
%The two states are related by the action of
%$ \CO_{z}  = e^{ -2 W_{\rm probe}(z) } \psi_Z^\dagger$ .
There exists \atish\ a BPS tunneling trajectory: \eqn\bpseqn{ 0 =
\delta \psi = \dot{ z}_{cl} - W_{\rm probe}^\prime(z_{cl}). } This
trajectory breaks just one supersymmetry, and thus supports a
single fermion zero mode.

We propose to identify the vacuum of the matrix model with fermion
number $k$ with the string theory vacuum with $k$ units of flux,
as defined in \S3. This identification is directly supported by
the interpretation of the MP matrix model as the worldline theory
of the unstable branes and of the D-instantons with the tunnelling
trajectories of the worldline tachyon field $Y$. Recall that an
unstable type II Dp-brane couples to the RR $p$-form potential
$C_{(p)}$ via the gradient of its tachyon according to $ \int
C_{(p)} \wedge d{\cal W}(Y) $ \refs{\HoravaJY,\SenMG,
\Billo,\Kennedy} with $V(Y) = {\del \over \del Y} {\cal W}(Y) $
\kutasovniarchos. For the unstable D-particle, this takes the form
\eqn\WZterm{ S_{\hat{D0}} \supset \int \chi ~{ dY \over dt} ~V(Y)
~dt. }
 Combined with our proposal,
this coupling implies that the tunnelling trajectory sources the
RR axion. It can therefore be identified with a D-instanton,
further vindicating the prescient analysis of \ShenkerUF.
%The operator
%$\CO$ can then be interpreted as the matrix model description of
%the D-instanton creation operator.

\newsec{Concluding remarks}

We have collected evidence supporting the conjecture that the
supersymmetric matrix model of Marinari and Parisi can be
identified with the matrix mechanics of $N$ unstable D-particles
in two-dimensional IIB string theory. This suggests that in a
suitable double scaling limit, the MP model,
when viewed from this perspective, provides a
non-perturbative definition of the string theory. The two systems
on both sides of the conjectured duality, however, clearly need
further study. We end with some concluding comments.

\subsubsec{Space-time fields}

An important open problem is the proper identification of the
space-time fields in the MP model. It is reasonable to expect
that, as for the $c=1$ \DJ\ and $\hat c =1$ cases, the spacetime
fields arise from the matrix model via the collective fields for
the eigenvalue density. Possibly the supersymmetric collective
field theory of
\refs{\JevickiYK,\CohnZJ,\vanTonderVC,\RodriguesBY} is the
correct framework, though it seems that some $Z_2$ projection may
be needed, since the target space boson $\chi$ and fermions
$\Upsilon$ are chiral with opposite chirality.
%\eqn\collective{
%\varphi(x) = \sum_i \ln (x - z_i),~~~
%\phi(x) \equiv \del_x \varphi(x), ~~~
%\psi(x) = \sum_i \psi_i {1 \over x - z_i}.
%}
If $\Upsilon$ is linear in the fermionic component of the
eigenvalue density (\eg\ a Laplace transform of it), then the
matrix model action of $(-1)^{F_L}: \psi \leftrightarrow \bar
\psi$, is consistent with the action on closed-string fields
$\Upsilon \leftrightarrow \bar \Upsilon$.

\subsubsec{Space-time supersymmetry}

Space-time supersymmetry should provide a helpful guideline in
finding a precise dictionary. Expanded around the quadratic
maximum of its bosonic potential, the model \quapot\ has many
symmetries; indeed the small bosonic and fermionic fluctuations
are decoupled. It is not difficult to find fermionic operators
which behave as in \pssbaralg. The more mysterious question is how
to describe the matrix model supersymmetry in the target space of
the string theory.

%hat the transformations \eqn\candidatesusy{ \delta_\epsilon \chi =
%\epsilon \Upsilon, ~~
% \delta_{\bar \epsilon} \chi = \bar \epsilon \Upsilon^\dagger, ~~
% \delta_{\bar \epsilon} \Upsilon = \del_t \chi + i F , \cdots}
%respect the interactions of $\phi$ does not seem entirely
%implausible.

\subsubsec{D-brane decay}

It should be possible to generalize the analysis
of  \refs{\McGreevyKB,\KlebanovKM,\McGreevyEP} to study the decay
of a single unstable D-brane. This will presumably involve a
superfield version of the fermion operator that creates and
destroys the super-eigenvalues, and a superfield bosonization
formula along the lines of \MartinecBG. This analysis would allow
an independent determination of the compactification radius of the
axion, along the lines of \refs{\KlebanovKM,\DouglasUP},
confirming that $\chi$ is periodic at the free fermion radius.

\bigskip
\noindent{\bf Acknowledgements}

JM would like to thank
Davide Gaiotto,
Emil Martinec, Leonardo Rastelli and Erik Verlinde
for conversations,
and
Dan Brace for discussions of supermultiplets of D-branes.
SM would like to thank
Chris Beasley,
Michal Fabinger, Mukund Rangamani,
Natalia Saulina and Nathan Seiberg
for useful conversations.
The work of JM is supported by a
Princeton University Dicke Fellowship.
This work is supported by the National Science
Foundation under Grant No. 98-02484.
Any opinions, findings, and conclusions or recommendations expressed in
this material are those of the authors and do not necessarily reflect
the views of the National Science Foundation.

\def\VV{{\cal V}}
\def\CA{{\cal A}}

\appendix{A}{The gauged Marinari-Parisi model}

In this appendix, we describe two ways of introducing an auxiliary
gauge field for the supersymmetric matrix model. We show that the
second method  is equivalent to the eigenvalue reduction of the MP
model given in \atish.

\bigskip

\noindent {\it Gauged Model, Version I}

The conventional method of gauging a supersymmetric action is to
introduce a real matrix superfield $\VV$, and replace the
%\be \VV = F_V + $\bar \theta \zeta + \bar
%\zeta \theta + \theta \bar \theta A.\ee
superderivatives in \act\ with gauge-covariant superderivatives
of the form \eqn\covone{ D_{{}_{\! \VV}} \Phi = e^{{\rm ad}\VV} D
(e^{-{\rm ad}\VV}\Phi) \, , \qquad \quad D =
\partial_{\theta} + \bar\theta \partial_\tau}\eqn\covtwo{\bar{D}_{{}_{\! \VV}} \Phi = e^{{\rm ad}\VV} \bar{D}_\theta
(e^{-{\rm ad}\VV}\Phi) \, \qquad \quad \bar{D} =
\partial_{\bar\theta} + \theta
\partial_\tau} These derivatives are designed to be covariant
under local gauge transformations \eqn\gaugeone{ \Phi \mapsto
e^{{\rm ad} \Lambda }~ \Phi \, , \qquad \quad \VV \mapsto \VV +
\Lambda} where $\Lambda$ is an arbitrary real superfield. We can
thus choose the gauge $\VV=0$. The presence of the gauge field
still manifests itself, however, by means of the requirement that
physical states must be annihilated by the generator of
infinitesimal {\it bosonic} gauge rotations $\Phi \mapsto
U^\dagger \Phi U$. This invariance can be used to diagonalize,
say, the bosonic component $\phi$ of the matrix superfield. The
off-diagonal fermionic matrix elements, however, remain as
physical degrees of freedom.

\bigskip

\noindent {\it Gauged Model, Version II}

A second possible procedure is to introduce a complex superfield
$\CA$ and define superderivatives \eqn\cova{ D_{{}_{\!\! \CA}}
\Phi = D\Phi - [\CA,\Phi]\, , \qquad \quad \bar{D}_{{}_{\!\! \CA}}
\Phi = \bar{D}\Phi - [\bar{\CA},\Phi] \, } covariant under local
gauge transformations \eqn\gaugetwo{ \Phi \mapsto e^{{\rm ad}
\Lambda }~ \Phi \, , \qquad \quad \CA \mapsto \CA + D\Lambda \, ,
\qquad \quad \bar{\CA} \mapsto \bar{\CA} + \bar{D}\Lambda} with
$\Lambda$ an arbitrary real matrix superfield. In this case, the
gauge invariance is not sufficient to choose a gauge in which
$\CA$ and $\bar\CA$ are set equal to zero. However, since $\CA$
and $\bar\CA$ appear as non-dynamical fields, we can eliminate
them via their equations of motion \eqn\gconst{ {\cal G} \equiv
[\, \Phi,\Pi\, ] =0 \, , \qquad \qquad \Pi \equiv D_{{}_{\!\!
\CA}} \Phi\, .} We will impose the physical state conditions in
the weak form \eqn\physc{{\cal G}\, | \Psi_{\rm phys}\rangle \,
=\, 0 \, .} The space of solutions to this constraint is
characterized as follows. Let $U$ be the bosonic unitary matrix
that diagonalizes the bosonic component $\phi$ of the matrix
superfield. We can then define \eqn\uphiu {(U\Phi U^\dagger)_{kk}
= z_k + \bar\theta\,\psi_k + \psi_k^\dagger\,\theta +
 \bar\theta\theta f_k \, . % \delta_{ij}.
} Here the $z_k$ are the eigenvalues of $\phi$. A straightforward
calculations shows that the gauge invariance conditions is solved
by physical states that depend on $z_k$ and $\psi_k$ only. Since
the physical state constraint \gconst\ is a supermultiplet of
constraints, this guarantees that this subspace forms a consistent
supersymmetric truncation of the full matrix model. One can also
verify directly that it is invariant under supersymmetry
transformations. After absorbing a factor  of $\Delta =
\prod_{i<j} (z_i-z_j)$ into our wave functions \eqn\split{
\Psi(z,\psi) = \Delta(z) \tilde{\Psi}(z,\psi)} the  supersymmetry
generators take the form \hermit.

\def\bra#1{{\langle}#1|}

\def\ket#1{|#1\rangle}

\def\downvac{|{\downarrow\downarrow\cdots\downarrow}\rangle}

It is convenient to think of the system of eigenvalues as $N$
particles moving in one dimension, each with an internal spin
$1\over 2$ degree of freedom, a spin ``up" or ``down.''
Accordingly, we can define the Hilbert space on which the
fermionic eigenvalues act by \eqn\downdowndown{ \psi_i \downvac =
0, ~~\forall i; ~~~ \psi^\dagger_i \downvac\equiv \ket{
\underbrace{\downarrow\cdots\downarrow}_{i-1}
\uparrow\downarrow\cdots \downarrow}  ; ~~~ etc... } A general
state in the physical Hilbert space is then
%%%%**THIS IS A STATE OF $\Psi_{new}$***
\eqn\generalstate{ \ket{f_{{\bf \eta}}} = \sum_{ {\bf \eta} }
f_{{\bf\eta}} ({\bf z}) \ket{\vec \eta} } where ${\bf \eta}$ is a
vector of $N$ up or down arrows, and we have arranged the
eigenvalues in a vector $\bf z$. Since it is possible via
$U(N)$-rotations to interchange any eigenvalue superfield
$(z_{i},\psi_{i})$ with any other eigenvalue superfield
$(z_{j},\psi_{j})$, the matrix wavefunctions should be symmetric
under this exchange operation.

However, since our wave-functions depend on anti-commuting
variables, the model will inevitably contain bosonic as well as
fermionic sectors. Let us now define a a fermionic interchange
operation $\kappa_{ij}$ with the property that it interchanges the
$i$ and $j$ spin state, and also multiplies the overall
wavefunction by a minus sign in case both spins point in the
up-direction. This minus sign reflects the Fermi statistics of
$\psi_{i}$ and $\psi_{j}$. Define the total exchange operation as
the product of the bosonic and fermionic one ${\cal K}_{ij} =
K_{ij}\, \kappa_{ij}$ where $K_{ij}$ interchanges $z_i$ and $z_j$.
We now specify the overall statistics of the physical
wavefunctions by means of the requirement that
\eqn\statistics{{\cal K}_{ij}\, |\tilde{\Psi}_{\rm phys}\rangle =
-|\tilde{\Psi}_{\rm phys}\rangle \, \qquad \qquad {\rm for\ all} \
{i,j}} The minus sign on the right-hand side ensures that the
original wavefunction, before splitting off the Vandermonde
determinant (see eqn \split), is completely symmetric. The
condition \statistics\ implies that particles with spin ``up" are
fermions, while particles with spin ``down" are bosons. We can
call this the spin-statistics theorem for our model.

%The standard quantum postulate implies the superspace commutation
%relations \be [\, \Pi(\theta_1),\Phi(\theta_2)\, ] = \theta_{12}
%\, , \qquad \quad [\, \bar\Pi(\theta_1),\Phi(\theta_2)\, ] =
%\bar\theta_{12}\, , \qquad \quad \{\,
%\Pi(\theta_1),\bar\Pi(\theta_2)\, \} = 1 \ee
%$$
%\Omega = U\, (1+ \theta \Upsilon + \bar\theta \Upsilon^\dagger
%+ {1\over 2}\theta \bar\theta
%[\Upsilon^\dagger,\Upsilon])
%$$
%$$\psi = {\rm diag}( \chi_1, \chi_2, \ldots,\chi_N)
%\left(\begin{array}{cccc}\psi_1 & 0 & \ldots & 0 \cr 0 & \psi_2 & \ldots & 0\cr
%0 & 0 & \ddots & 0

\appendix{B}{Boundary states for $\CN=2$ Liouville}

In this appendix,
we will attempt to write down
the boundary state
for the unstable D-particle
of type IIB in the $\CN=2$ Liouville
background.
In doing this, we will take advantage of the
worldsheet $\CN=2$ supersymmetry
by expanding the boundary
state in Ishibashi states
which respect the $\CN=2$.
%As we will review, such states come in
%two varieties, A and B, see \eg\ \OoguriCK.
%BPS D-branes in type IIA/B
%can be expanded in A/B-type Ishibashi states,
%which are therefore SCFT descriptions
%of supersymmetric cycles.
%On the other hand, in type IIB,
%an A-type $\CN=2$ boundary state
%describes an unstable brane (or brane-antibrane pair)
%on the same
%supersymmetric cycle.
%The difference is that
%it is not possible to write
%a corresponding GSO-invariant RR piece of the boundary state
%with the same wavefunction.
%In this appendix therefore, we are describing
%non-BPS branes on supersymmetric cycles.
In order to write the Ishibashi states,
we will need to recall some facts about the
primaries on which they are built, and their
characters.

\subsubsec{Characters of the $\CN=2$ superconformal algebra}

The chiral $ \CN = 2 $ characters are defined by
\eqn\ellipticgenus{
 \chi_{V} (q, y) =
{\rm Tr}_{V} q^{ L_0 - c/24 } y^{J_0} .
}
The trace is over an $\CN = 2$ module $V$.
These representations are built on
primary states
labelled by the eigenvalues $h,\omega$ of the
central zeromodes $L_0, J_0$.
%We call the R-charge $\omega$
%because it plays the role of energy
%in this system.
It will be convenient to label
our primary states
by $ P$ and  $\omega$,
related to the conformal dimension
by $ h = ( Q^2/4 + P^2 + \omega^2 )/2 = ( 1 + P^2 + \omega^2)/2$.
The Liouville momentum $P$\foot{
Here we are defining Liouville momentum as $P$
appearing in the wavefunction
$ e^{ -( Q/2 + i P) \rho}$.}
is determined by this equation
up to choice of branch, both of which have
the same character.

These characters were written down in \BoucherBH.
For the module associated with a generic
NS primary, labelled by $[P,\omega]$, the
character is
\eqn\NScharacter{
\chi^{NS}_{[P,\omega]}(q, y) =
%q^{h - (c-3)/24}
q^{ P^2 /2 + \omega^2 /2} ~
y^\omega { \vartheta_{00}(q,y) \over \eta^3(q)}
}
In the R-sector, a primary
is also annihilated by $G^+_0$ or $G^-_0$,
and this results in an extra label $\sigma = \pm$
on the character.
For the primary with labels $P,\omega,\sigma$,
the character is (let $y \equiv e^{2 \pi i \nu}$)
\eqn\Rcharacter{
\chi^R_{[P,\omega, \sigma]}(q,y) =
2 \cos \pi \nu~
%q^{h - (c-3)/24}
q^{ P^2/2 + \omega^2/2} ~
~y^{\omega + {\sigma\over 2}}
{ \vartheta_{1,0}(q,y) \over \eta^3(q)}
}
We will also need the character
for the identity representation
\eqn\idcharacter{
\chi^{NS}_{{\bf 1}}(q,y) = q^{-1}{ 1 - q \over ( 1 + y q^{1/2} )
( 1 + y^{-1} q^{1/2})} { \vartheta_{0,0}(q,y) \over \eta^3 (q)} .
}
%For degenerate representations
%of the algebra, the characters are different,
%but can be written as
%linear combinations of the nondegenerate
%characters.
%Under spectral flow by an amount
%$\eta$,
%the characters transform to
%\eqn\spectralflowbyeta{
%\chi^\eta(q, y)
%= q^{ c \eta^2 } y^{2c\eta} \chi( q, yq^{\eta}).}
%By spectral flow by a half unit
%from the NS sector, we reach the R sector:
%\eqn\spectralflowtoR{
%\chi^R_{1}(q,y) \equiv q^{c/4} y^{2c}
%\chi_1^{NS}(q,yq^{1/2}).
%}

\subsubsec{Modular properties of the characters}

The modular transformation properties
of the chiral characters
of the $\CN=2$ algebra will be crucial
for our study of D-branes
in $\CN=2$ superLiouville.
We use the notation
$ \tilde q = e^{2 \pi i \tau}, ~
\tilde y = e^{ 2 \pi i \nu}$
for closed string modular variables,
and $q = e^{ - 2 \pi i / \tau}, ~ y = e^{ \pi i \nu / \tau}$
for their open string transforms.
%Some relevant properties of theta
%functions have been collected AT THE END.
The characters
%\NScharacter\Rcharacter\idcharacter\
participate in the following formulas \Ahn\Eguchi.
%\eqn\nondegenNS{
%\int_{- \infty}^\infty d\tilde \omega ~ e^{- 2 \pi i\omega \tilde \omega}
%\int_{-\infty}^\infty d\tilde P ~e ^{2 \pi iP \tilde P}~
% \chi^{NS}_{[\tilde P,\tilde \omega]}(\tilde q, \tilde y)
%= \chi^{NS}_{[P, \omega]}(q, y)
%}
%\eqn\nondegenR{
%\int_{- \infty}^\infty d\tilde\omega
%~ e^{ -2 \pi i\omega \tilde \omega}
%\int_{-\infty}^\infty d\tilde P ~e ^{2 \pi iP \tilde P}~
% \chi^{NS}_{[\tilde P,\tilde \omega]}(\tilde q, - \tilde y)
%={ 1 \over \cos \pi \nu} ~ \chi^{R}_{[P, \omega]}(q, y).
%}
%\eqn\degenNSagain{
%\int_{-\infty} ^\infty d \tilde P
%d\tilde \omega~
%~e^{- 2 \pi i \omega \tilde \omega}
%~2 \sinh^2 \pi \tilde P ~ \chi^{NS}_{[\tilde p, \tilde \omega]}
%( \tilde q, \tilde y)
%=
%y^{ \omega} ~ q^{ \omega^2 /2}~
%\left( q^{ - 1/2} - 1\right)
% { \vartheta_{00}(q,y) \over \eta^3 (q)}
%}
\eqn\degenNSagain{
\int_{-\infty} ^\infty d \tilde P
d\tilde \omega~
~S(\tilde p, \tilde \omega)
~ \chi^{NS}_{[\tilde p, \tilde \omega]}
( \tilde q, \tilde y)
=
\chi_{{\bf 1}}^{NS}(q,y)
}
\eqn\degenRagain{
\int_{-\infty} ^\infty d \tilde P
d\tilde \omega~
~S(\tilde p, \tilde \omega)
~ \chi^{NS}_{[\tilde p, \tilde \omega]}( \tilde q, -\tilde y)
\equiv
~ \chi^{R}_{{\bf 1}}(q,y)
}
\eqn\NSmodularmatrix{
S(\tilde p, \tilde \omega) =
{ \sinh^2  \pi \tilde p \over
2 \cosh( \pi p /2 + i \pi \omega/2) \cosh( \pi p/2 - i \pi \omega/2) }.
}
%
%\eqn\Rmodularmatrix{
%S_{R}(\tilde p, \tilde \omega) =
%{ \sinh^2  \pi \tilde p \over
%2 \sinh( \pi p /2 + i \pi \omega/2) \sinh( \pi p/2 - i \pi \omega/2) }
%.
%}
%\eqn\degenRagain{
%\int_{-\infty} ^\infty d \tilde P
%d\tilde \omega~
%~e^{ -2 \pi i \omega \tilde \omega}
%~2 \sinh^2 \pi \tilde P ~ \chi^{NS}_{[\tilde p, \tilde \omega]}
%( \tilde q, -\tilde y)
%=
%y^{ \omega \pm 1/2 } ~ q^{  (  \omega \pm 1/2)^2 /2}~
%\left( q^{ - 1/2} - 1\right)
% { \vartheta_{10}(q,y) \over \eta^3 (q)}
%}
%The objects on the RHS of \degenNSagain\degenRagain\
%can be written in terms of characters
%of degenerate representations, if desired.
Note that these formulas \degenNSagain--\NSmodularmatrix\ are
relevant for the case that
R-charge is not quantized,
on which we focus for simplicity
in our study of boundary states.
The refinement of these formulas
to the case of compact euclidean time
follows from \degenNSagain--\degenRagain\
by Fourier decomposition.
Further, these formulas arise
by performing a formal sum;
a careful treatment
of convergence issues
reveals additional contributions
from discrete states \Eguchi.
%is
%In the case when the R-charge is quantized
%(meaning for us that the Euclidean time direction is
%compact), the following formulas are useful
%\eqn\modularzerozero{
%\sum_{n \in \IZ} e^{2 \pi i n r }
%\int _{- \infty}^\infty d\tilde P~
%e^{ 2 \pi i P \tilde P}
%%\chi^{NS}_{[h(P,n),n]} (\tilde q, \tilde y)
%\chi^{NS}_{[\tilde P,n]} (\tilde q, \tilde y)
%= { 1 \over \sqrt 2}
%\sum_{m \in \IZ} \chi^{NS}_{[P,m-r]}(q,y)
%}
%\eqn\modularzeroone{
%\sum_{n \in \IZ} e^{2 \pi i n r}
%\int _{- \infty}^\infty d\tilde P~
%e^{  2 \pi iP \tilde P}
%\chi^{NS}_{[\tilde P,n]} (\tilde q,- \tilde y)
%= { 1 \over \sqrt 2}
%\sum_{m \in \IZ} \chi^{R}_{[\tilde P, m-r]}(q,y)
%}
%\eqn\modularzerozerodegen{
%\sum_{n \in \IZ} e^{2 \pi i n r }
%\int _{- \infty}^\infty d\tilde P~
%2 \sinh^2 \pi \tilde P ~
%%\chi^{NS}_{[h(P,n),n]} (\tilde q, \tilde y)
%\chi^{NS}_{[\tilde P,n]} (\tilde q, \tilde y)
%=
%%{ 1 \over \sqrt 2}
%\sum_{m \in \IZ}
%y^{ - 2 (m -r) } ~ q^{  (m-r) ^2 /2}~
%\left( q^{ - 1/2} - 1\right)
% { \vartheta_{00}(q,y) \over \eta^3 (q)}
%%\chi^{NS}_{[P,m-r]}(q,y)
%}
%\eqn\modularzeroonedegen{
%\sum_{n \in \IZ} e^{2 \pi i n r }
%\int _{- \infty}^\infty d\tilde P~
%2 \sinh^2 \pi \tilde P ~
%\chi^{NS}_{[\tilde P,n]} (\tilde q, - \tilde y)
%=
%%{ 1 \over \sqrt 2}
%\sum_{m \in \IZ}
%y^{ - 2 (m-r) - 1 } ~ q^{  (  m  -r + 1/2)^2 /2}~
%\left( q^{ - 1/2} - 1\right)
% { \vartheta_{10}(q,y) \over \eta^3 (q)}
%%\chi^{NS}_{[P,m-r]}(q,y)
%}

\subsubsec{Ishibashi states for $\CN=2$}

Using this
notation for
representations of the $\CN=2$ algebra, let
us now study Ishibashi states
based on these representations\foot{Useful references
include \refs{\OoguriCK,\GaberdielJR,\BDLR,\SenBdy}.}.
Such states
\IshibashiKG\
provide a basis
for D-brane states
which respect the
$\CN=2 $ algebra.
They carry two kinds of labels:
those which specify
the primary of
the chiral algebra on which
the state is built;
and those which specify
the automorphism of the chiral
algebra which was used to
glue the left and right chiral
algebras.  In our notation, we will
separate these labels by a semicolon.
Only automorphisms of the $\CN=2$
which preserve the gauged
$\CN=1$ subalgebra are allowed.

There is a $\IZ_2$ automorphism group of the
$\CN=1$ subalgebra ($ G = G^+ + G^-$)
\eqn\etamap{G \to \eta G}
with $\eta = \pm 1$.
There is an additional $\IZ_2$ automorphism
of the $\CN=2$ algebra,
which commutes with \etamap, and which is
generated by
$$G ^\pm \to G^{\pm \xi}, J \to - \xi J $$
with $\xi = \pm$
\OoguriCK\
(the trivial map, $\xi = +1 $ is B-type,
the nontrivial map $\xi = -1$ is A-type).
%I don't think there exist
%other automorphisms of the $\CN=2$ which
%are involutive on the gauged subalgebra.

Let $j$ label the $\CN=2$ primaries; it
is a multi-index
with three components:
$$ j = \left[ h, n, \left\{ \matrix{ NS \cr R+ \cr R-}
\right\} \right] .$$
(recall that an R primary is
further specified by whether it
is annihilated by $G^+_0$ or $G^-_0$.)
%A B-type Ishibashi state satisfies
%\eqn\btypeishi{
%( L_n - \tilde L_n ) \kket{ j; B \eta}, ~~~
%( G^\pm_r - i \eta \tilde G ^\pm_r )
%\kket { j;B \eta} = 0 , ~~~
%( J_n - \tilde J_n) \kket{ j;B \eta}
%}
An A-type Ishibashi state satisfies
\eqn\btypeishi{
( L_n - \tilde L_n ) \kket{ j; A,\eta}, ~~~
( G^\pm_r - i \eta \tilde G ^\mp_r )
\kket { j;A, \eta} = 0 , ~~~
( J_n - \tilde J_n) \kket{ j;A, \eta}
}
where
$r$ is half-integer moded
if $\ket {j}$ is an NS primary,
and integer moded if $\ket{j}$
is from an R sector.

In order to make type II D-branes
from these states, we will need to
know the action of the fermion number
operators on them.
In the NS sector \GaberdielJR,
\eqn\fermionnumberonNS{
(-1)^F \kket{ j, NS; \xi, \eta} = - \kket{ j, NS; \xi, - \eta} =
(-1)^{\tilde F} \kket{ j, NS; \xi, \eta}.
}
Note that the existence of
the state with one value of $\eta$
plus the conserved chiral fermion number
implies the existence of the other.
In the R sector, the action is more subtle because
of the fermion zero modes.
Since unstable branes may be built without using
RR Ishibashi states, we will not discuss them further.

%For A-type boundary states, the
%reflection conditions on the supercurrents
%and R-currents are replaced by
%$$
%( G^\pm_r - i \eta \tilde G ^\mp_r )
%\kket { j; A\eta} = 0
%$$
%$$
%( J_n + \tilde J_n) \kket{ j;A \eta}.
%$$

Next we need to know the
matrix of inner products between these
states.  The inner product
is defined by closed-string
propagation between the
two ends of a cylinder:
%Therefore, it depends on the
%GSO projection which
%is being performed.
%We are going to suppose that we can ignore this
%subtelty
%(at least for B-type boundary states)
%[this something we can understand in finite time],
%in which case we get
\eqn\closedannulusishibashi{
\bbra{ j_1; \xi_1\eta_1} ~D(\tilde q, \tilde y)
~\kket{ j_2; \xi_2\eta_2}
= \delta( j_1, j_2)~
\delta_{\xi_1,\xi_2} ~ \chi_{j_1} ( \tilde q, \eta_1\eta_2
\tilde y)
}
where $D(\tilde q, \tilde y)$ is the
%properly GSO-projected
closed-string propagator,
twisted by the R-current.
%\foot{
%This formula should be
%regarded with some suspicion%%%
%
%because of the non-diagonal closed-string
%propagator.]
The delta-function on primaries is
obtained from the overlap of closed-string states:
\eqn\deltafunction{
\delta(j_1, j_2) \equiv
\left( \bra{ j_1 } \otimes \bra{ \tilde{j_1}} \right)
\ket{j_2} \otimes \ket{ \tilde j_2}
}

\subsubsec{The D0 boundary state}

%Experience with
%Liouville models with
%less worldsheet supersymmetry
%suggests that to look for branes
%localized in the Liouville direction,
%we should study boundary states
%associated with
%the (NS) vacuum state,
%which we will denote $\ket{(1,1)}$,
%after the related object in the bosonic
%Liouville and $\CN=1$ superLiouville theory.

We would now like to describe 
the boundary state for the unstable D-particle.
Since it is extended in the R-symmetry direction,
it should be a B-type brane.
To construct consistent boundary
states for $\CN=2$ Liouville,
we will follow the strategy
which was successful for
bosonic Liouville \refs{\ZZB,\FZZ,\TLbound},
and for $\CN=1$ Liouville \refs{\AhnEV,\FukudaBV}.
Basically,
the consistent boundary states are Cardy states
\TLbound;
their
wavefunctions can be
written in terms of the modular matrix
$ U_j(i) = {S_j^i  \over \sqrt {S_0^i}}$
\CardyIR.
To be more precise, suppose,
as in \ZZB,
that we can expand
the desired boundary state
in
Ishibashi states for the non-degenerate
(NS)
representations of the $\CN=2$ algebra:
\eqn\ZZhypothesis{
\ket{ B_0, \eta} =
\int_{-\infty}^\infty d\omega \int_{-\infty}^\infty dP
%\sum_{n= - \infty}^{\infty}
~U^\eta(P, \omega)
~\kket{P, \omega, NS; B,\eta}.
}
%is independent of $\eta $,
%as it is for the D-particle in
%$\CN=1$ Liouville \AhnEV\FukudaBV.
In writing an integral over $\omega$
in \ZZhypothesis,
we are focusing on the case when
the time direction is infinite in extent.
To build non-BPS type II branes from
these boundary states,
GSO-invariance requires
$U(P,\omega) = U^+(P,\omega) = - U^-(P,\omega)$
according to \fermionnumberonNS.
We will therefore write
%
%The general formula \unstable\
%makes
%the basic GSO-invariant boundary state
%for an
%unstable D0-brane of type IIB
\eqn\generalform{
\ket{ \hat{D0}}
=
%{1 \over \sqrt 2}
%\left(
\kket{ B_0, \eta = +1 }- \kket {B_0 , \eta = -1} .
%\right).
}
%Note that the prefactor of $\sqrt 2 $
%corrects the boundary entropy
%so that this represents a single brane.

Next, we make the self-consistent
assumption that
%this brane exists, and
%that
the bosonic open-string spectrum
should contain
only
states with Liouville momentum
$P = -i $
corresponding to the identity state.
Given \degenNSagain\
a solution to this requirement
is \Ahn\Eguchi
\eqn\wavefunction{
U(P, \omega) = \sqrt 2 \cdot e^{i \delta(P,\omega)}\cdot
{ \sinh \pi P \over \cosh ( \pi p/2 + i \pi \omega/2) }.
}
%This Cardy-type ansatz for the
%boundary state wavefunction
%has proved extremely successful for
%less-supersymmetric Liouville theories,
%The phase $e^{i \delta(P,\omega) } \equiv
%\mu^{ iP } { \Gamma( 1 + i P ) \over \Gamma( 1 - iP) } $
%is analogous to the standard leg-pole factor,
The phase $e^{i \delta(P,\omega) } $
is not actually determined by the
modular hypothesis.
As we will verify next,
the D-brane with this wavefunction \wavefunction\
indeed has only the identity representation
in its bosonic open-string spectrum.

%This means that it describes
%an object which is pointlike in both time and space,
%an instanton.  More specifically, in type IIB, it
%describes a D-instanton anti-D-instanton pair.

%A wavefunction that describes a D-particle
%and satisfies our requirement is
%\eqn\particlewavefunction{
%\hat{U}(P,\omega) = \sqrt 2
%\cdot e^{ i \delta(P)} \cdot \sinh \pi P ~ \delta (\omega).
%}
%We will also study the
%rolling tachyon wavefunction
%\eqn\rollingwave{
%{U}_{\rm rolling} (P,\omega) = \pi
%{ \sinh \pi P \over \sinh \pi \omega} .}
%This one-point
%function should be checked by bootstrap methods.

\subsubsec{Open string spectrum}
%The unstable D0 of type IIB}

The vacuum annulus amplitude between the
boundary state
and itself,
\eqn\annulusDzero{
\CA(q) = \bra{\hat{D0}} D(\tilde q) \ket{\hat{D0}} ,
}
determines the open string spectrum by channel duality.
In this expresseion, $D(\tilde q)$ is
the closed-string propagator, on a
tube of length $\tau = \ln \tilde q / 2 \pi i $.
Given the expression
\ZZhypothesis,
it takes the form
%\eqn\annulusDzerozero{
%\eqalign{
%&\CA(q) = \int P_1 \int dP_2 \int dr_1 \int dr_2
%~ \sum_{\eta_1, \eta_2= \pm} \eta_1 \eta_2  ~\times \cr
% &U(P_1, r_1) U^\dagger(P_2, r_2) ~ \times
%\bbra{ P_2,r_2, NS ; B \eta_2} D(\tilde q)
%\kket{P_1, r_1, NS; B \eta_1}.
%}
%}
\eqn\annulusDzeroone{
\CA(q) =
\int_{-\infty}^\infty dP~d\omega~
\hat U(P,r) ~\hat U^\dagger(P,\omega)
\sum_{\eta_1 \eta_2} \eta_1 \eta_2 ~
\chi^{NS}_{[P,\omega]}(\tilde q, \eta_1 \eta_2)~
\chi^{NS}_{gh}(q, \eta_1,\eta_2).
}
%
%\eqn\annulusDzeroone{
%\eqalign{&\CA(q) = \int dP~\int dr~ U(P,r) ~U^\dagger(P,r)
%\sum_{\eta_1 \eta_2} \eta_1 \eta_2
%\times \cr & \left(
%\chi^{NS}_{[P, r]}(\tilde q, +1)
%+
%\chi^{NS}_{P, r}(\tilde q, +1)
%-
% \chi^{NS}_{P, r}(\tilde q, -1)
%-
%\chi^{NS}_{P, r}(\tilde q, -1)
%\right)
%}}
Note that there are two sets
of terms which are identical.
This reflects the fact that unstable branes
are $\sqrt 2 $ times heavier than BPS D-branes.

Using the modular transformation formulas above,
and the wavefunction \wavefunction,
in the open-string channel this reads
\eqn\annulusopen{
\CA(q) = 2 \left( \chi_1^{NS}(q,1)~\chi_{gh}^{NS}(q,1)
+ \chi_1^R(q,1)~\chi_{gh}^R(q,1) \right).
}
%THIS FORMULA NEEDS TO BE IMPROVED,
%\eqn\annulusopen{
%\CA(q) = V_t~
%\int_{-\infty}^\infty d\omega ~
%\left( q^{-1/2} - 1 \right)
%\left(y^\omega q^{ \omega^2 /2}
%{ \vartheta_{00}(q,y) \over \eta^3(q)}
%- y^{ \omega + 1/2} q ^{ ( \omega + 1/2)^2/2}
%{ \vartheta_{10}(q,y) \over \eta^3(q)} \right).
%}
%$V_t = \delta(0) $ is the volume of the time
%direction.
%In terms of characters this may be written
%\eqn\annulusopenincharacters{
%\CA(q) =
%V_t \int d\omega ~q^{ - \omega^2 /2}
%\left( \chi^{NS}_{(1,1)} (q)
%- \chi^R_{(1,0)}(q)
%\left( \chi^{NS}_{[0,1]} (q,1)
%- \chi^R_{[0,1/2,+]}(q,1)
%\right).
%}
%\foot{Note that really what we mean by $\chi^R$ here is
%$ \half (\chi^R_{1+} + \chi^R_{1-} )$
%which contributes the fermions $\psi$ and $\psi^\dagger$
%respectively.
%}
As usual for two-dimensional
strings,
the contribution from the
 $\CN=2$ $bc\beta\gamma$ ghost system
% has the effect of cancelling
cancels all of the modular functions from each term, and
we find
\eqn\finalspectrum{
Z \equiv \int {dt \over 2 t} \tr ~q^{H_{{\rm open}}}
= \int {dt \over 2 t}
~\left[
\left( q^{- 1/2} - 2 + \dots \right)
- \left( 1 + \dots \right)
\right]
}
with $ q \equiv e^{ - \pi t}$.
%\foot{
%BEWARE RELATIVE PREFACTORS OF 2]}
Deducing the mass of the states propagating in the open string loop is
subtle because the time direction participates
in the $\CN=2$ algebra, and the
analytic continuation needs to be understood better. We can however
understand the spectrum by looking at the large-$t$ behaviour of the amplitude.
The first term in brackets is the contribution 
of the NS sector.
The two contributions indicated give rise
to divergences and represent a tachyon $Y$, and a complex 
massless discrete
state, respectively. 
This is similar to the case of 
the unstable D-particle
in the bosonic $c=1$ string\foot{
upon multiplying $\alpha^\prime$ by 2}
(\qv\ the lovely appendix B of \KlebanovKM)
and in the $\hat c=1$ type 0B string
\refs{\TakayanagiSM,\DouglasUP}.

The second term in brackets is the
contribution of the R sector,
which produces the fermions which were absent
in the above examples.
The term indicated represents a massless
complex fermion $\psi$,
%and a massive fermionic discrete state $\xi$.
%Note that the mass splitting between
%the tachyon and the massless fermion
%is the same as the splitting between
%the bosonic discrete state and
%the fermionic discrete state.
%in accord with
%\spectralflowonboundary\spectralflowonboundaryferm.

This is the spectrum
of the gauged
Marinari-Parisi model \act\
on a circle.
The appearance
of non-quantized
energies at intermediate
stages can
be remedied \Eguchi\
by employing
characters of
the $\CN=2$ algebra
extended by the
spectral flow generators.

%For the rolling brane \rollingwave\
%following the analysis of \KarczmarekXM\
%gives the following expression for
%the annulus amplitude
%\eqn\rollingannulus{
%\CA_{\rm rolling} (q) =
%\int d \omega ~\rho(\omega) ~
%\left( \chi^{NS}_{[0, \omega]}(q,1)
%- \chi^{R}_{[0, \omega + 1/2]}(q,1) \right).
%%q^{ \omega^2/2} y^\omega
%}
%with
%an open-string density of states
%$ \rho(\omega) = \del_\omega \ln S_{1} ( 1- \omega)$
%related to the q-gamma function.

For the instanton-anti-instanton pair,
the only difference is that
we use A-type Ishibashi states,
which describe branes
which are localized
in the R-symmetry direction.
%The annulus
%only contains the identity representations (of the NS and R
%sectors) \AlexandrovNN\ in the open string channel:
%\eqn\instantonannulus{ \CA_{\rm instanton} (q) =
%\chi^{NS}_{1}(q,1) - \chi^{R}_{1}(q,1).}
The spectrum
of the BPS D-instanton is obtained from this by an open-string GSO
projection, and therefore has half as many fermion zeromodes,
namely one.

Clearly we have merely
begun to study
the interesting zoology of D-branes in this system.
Further development appears in \Ahn\Eguchi.

\listrefs
\end

\subsubsec{For our convenience, the $\CN=2$ algebra}

$$\eqalign{
    T(z)T(w) = { c/2 \over (z-w)^4} + {2\, T(w) \over (z-w)^2} +
         {\partial_w T(w) \over z-w} + \cdots ~,\cr
    T(z)G^{\pm}(w) = {3/2 \over (z-w)^2}\, G^\pm(w) +
           {\partial_w G^{\pm}(w) \over z-w} + \cdots ~,\cr
    T(z)J(w) = {J(w) \over (z-w)^2} + {\partial_w J(w) \over z-w} + \cdots~,\cr
    G^{+}(z)G^{-}(w) = {2\,c/3 \over (z-w)^3}
                      + {2J(w) \over (z-w)^2} +
               {2T(w) + \partial_w J(w) \over z-w} + \cdots~,\cr
    J(z)G^{\pm}(w) = \pm\, {G^{\pm}(w) \over z-w} + \cdots~,\cr
    J(z)J(w) =  {c/3\over (z-w)^2} + \cdots~.
}
$$

In modes,
$$
     J(z) = \sum_n\, J_n\, z^{-n-1} ~,~~
     G^{\pm}(z) = \sum_n\, G^{\pm}_{n \pm a}\, z^{-(n \pm a) - 3/2}~,~~
     T(z) = \sum_n\, L_n\, z^{-n-2}~.
$$
The parameter $a$ in the mode expansion is an integer for the R
sector (anti-periodic boundary conditions on the fermionic
currents) and a half-integer for the NS sector (periodic boundary
conditions on the fermionic currents). ( $    G^{\pm}(e^{2 \pi i}
z) = -     e^{\mp 2 \pi i a}\, G^{\pm}(z)~.$) [Note that it
commutes with the candidate $ H = L_0 + \tilde L_0 - p_R p_L $. ]

In the last part of this paper we study a rolling eigenvalue,
using an adiabatic description as in \McGreevyKB\McGreevyEP. A
probe super-eigenvalue $Z = z + \psi_Z \bar \theta + \cdots$ feels
the mean-field superpotential \probesuperpotential. Dually, since
\probesuperpotential\ depends on $X_i$, its presence modifies the
ground states of the already-present super-eigenvalues. If we put
the probe in its down state, the modification of the $k=0$ ground
state is \eqn\probedgroundstate{
%\ket{f_k} \to \ket{f_k ( z, \downarrow) }
%= N_k(z, \uparrow) ~e^{ -W_{\rm probe} (z) } \ket{f_k} .
\ket{f_0} \to \ket{f_0 ( z, \downarrow) } = N_0(z, \uparrow) ~e^{
-W_{\rm probe} (z) } \ket{f_0} . } As in \McGreevyEP, let us
compute the Berry phase resulting from this dependence on the
probe. The Berry connection transporting this state is
\eqn\berryconnection{
%A_z(k) = \bra{ f_k (z, \downarrow)}
%{\del \over \del z} \ket{ f_k (z, \downarrow)}
A^\downarrow_z(0) = \bra{ f_0 (z, \downarrow)} ~{\del \over \del
z} ~\ket{ f_0 (z, \downarrow)} = \half \del_z \ln |N|^2 } where
\eqn\macroscopicloop{
% \ln |N(z)|^{2} = - \ln \vev{ e^{ - 2 W_{\rm probe}(z) } }_k
%\simeq 2 \vev{ W_{\rm probe} (z) }_k .
 \ln |N(z)|^{2} = - \ln \vev{ e^{ - 2 W_{\rm probe}(z) } }_0
\simeq 2 \vev{ W_{\rm probe} (z) }_0. } The Berry phase acquired
by the state $\ket{f_0}$ when the probe super-eigenvalue executes
a trajectory $z(t)$ is then \eqn\berryphase{ \exp{\left( i
\int^\tau dt ~\dot z ~A^\downarrow_z(0) \right)} = e^{ i
\varphi(z(\tau))} . } Here we have used the fact that the
collective field is determined by the discontinuity of the probe
superpotential by the relation \collectiveexpansion. The creation
operator for this probe should therefore be \eqn\downprobecreator{
\hat \Psi(z(\tau), \downarrow) = \psi_0(z,\tau) \cdot e^{ i \hat
\varphi(z, \tau) }
%e^{ \hat \Phi(z, \tau)}
} where $\psi_0$ satisfies the single-particle Schrodinger
equation with Hamiltonian $\hat p^2 /2 + V(z) $. In terms of these
modes, the $\CN = 2$ superconformal algebra takes the form:
%%%%%%%%%%%%%%%%%%%%%%%%%%%%%%
$$
\eqalign{
     \left[ L_m, L_n\right] = (m - n) L_{m+n} +
             {c\over 12}\, m (m^2 - 1)
          \,\delta_{m+n,0}~,   \cr
     \left[ J_m, J_n\right]  ={c\over 3} m\, \delta_{m+n,0}~, \cr
     \left[ L_n, J_m\right]  = -m\, J_{m+n}~,  \cr
     \left[ L_n, G^{\pm}_{m \pm a}\right] =
              \left( {n\over 2} - (m \pm a)\right) \, G^{\pm}_{m + n\pm a}~, \cr
     \left[ J_n, G^{\pm}_{m \pm a}\right]  =\pm\, G^{\pm}_{m + n \pm a}~,\cr
    \left\{G^+_{n+a}, G^-_{m-a}\right\} = 2\, L_{m+n} +
      (n-m+2a)\, J_{n+m}
     \cr
  ~~~ +
      {c\over 3} \left( (n+a)^2 - {1 \over 4} \right)
      \delta_{m+n,0}~.
}$$
For this quantity one finds \eqn\imannulus{ \int_{1/\epsilon}  {dt
\over t}
%e^{ 2 \pi t L_0}
~{\rm Im} \left( \int d\nu ~\rho(\nu) ~ e^{  2\pi t \nu^2 }\right)
= \int_{1/\epsilon}  {dt \over t} ( q^{ 1/2} - 1 ) q^{ -1/2}
+\cdots = \ln \epsilon  + \cdots } where $\cdots$ indicates finite
terms, $\rho(\nu)$ is the open-string density of states for the
rolling tachyon \refs{\KarczmarekXM,\McGreevyEP}, and $q = e^{
2\pi /t}$. \subsubsec{And some relevant properties of theta
functions}

$$
q = e ^{2 \pi i \tau}, ~~~ \tilde q = e^{ 4 \pi^2 / \ln q} = e^{- 2 \pi i /\tau}
$$

[I SHOULD HAVE KEPT THE OTHER ARGUMENT HERE.  I'LL PUT
IT BACK LATER.]
$$
\eta(\tilde q) = \sqrt{ - {\ln q \over 2\pi}} \eta(q)
$$
$$
\vartheta_{00}(\tilde q ) = \sqrt{ - {\ln q \over 2\pi}} \vartheta_{00}(q)
$$
$$
\vartheta_{10}(\tilde q) = \sqrt{ - {\ln q \over 2\pi}} \vartheta_{01}(q)
$$
$$
\vartheta_{01}(\tilde q) = \sqrt{ - {\ln q \over 2\pi}} \vartheta_{10}(q)
$$
Note that
$ \vartheta_{00}  \equiv \vartheta_3$,
$\vartheta_{01} \equiv \vartheta_4$,
$ \vartheta_{10} \equiv \vartheta_2$.

Also,
$$ \vartheta_{00}(\nu + {1\over2}, \tau)
 = \vartheta_{10}(\nu, \tau).$$
and
$$ \vartheta_{00}(\nu - \tau/2, \tau)
= \sqrt y \vartheta_{01}(\nu, \tau) $$
[this last one needs to be checked again.]

\listrefs

\end

First, we review the gauge fixing of the $c=1$ matrix model,
following \refs{\BKZ, \GinspargIS, \igorreview}. Consider the
vacuum path integral for the gauged $c=1$ matrix model,
\eqn\bosonicpath{ Z(T) = \bra{0}e^{ - HT} \ket{0} = \int [ D\Phi]
[DA]~e^{ \int_0^T dt~{\cal L}[\Phi,A]}. } To proceed, it is useful
to change variables to radial coordinates by inserting in the path
integral \eqn\resolutionofone{ 1 = \int [d\Lambda] \int [d\Omega]
~\delta[ \Phi - \Omega \Lambda \Omega^{-1} ] {\cal J} } where
$\Lambda = {\rm diag}(\lambda_1, \dots,\lambda_N)$ and $\Omega$
are functions of $t$. The Jacobian for this transformation is
\eqn\localvandermonde{ {\cal J} = \exp{\left(
 2 \int dt \sum_{i < j} \ln \left( \lambda_i(t) - \lambda_j(t)
\right) \right)}. } The condition that the wavefunctions are gauge
invariant is \eqn\gaugeinvariant{ 0 = { \del\over \del \Omega}
\Psi_{new}( \Phi) . } Local gauge-fixing can then be accomplished
by setting $ D_t \Omega = 0 $, \ie\ $\Omega(t) = e^{ i \int^t A}
$. The path integral over $A$ then cancels all of the vandermondes
except the first and last \BKZ. It is then possible \BIPZ\ to
absorb these into the initial and final wavefunctions,
\eqn\newbosonicwavefn{ \Psi_{\rm new}(\Lambda, t) \equiv e^{ -
\sum_{i<j} \ln \left( \lambda_i(t) - \lambda_j(t) \right)}
\Psi_{\rm old} (\Lambda, t). }

\eqn\tachyon{V_Y= e^{-\varphi -2\rho} \, e^{i p_\theta \theta} \, \qquad \qquad p_\theta^2 =1}
\eqn\fermion{V_\Psi= e^{-{\varphi\over 2}  -\rho}\, e^{i{H/ 2}} \, \qquad \quad
V_{{\bar \Psi}}= e^{-{\varphi\over 2}  -\rho} \, e^{- i{H/ 2}} }
\eqn\afield{V_A= e^{-\varphi-2\rho} \, \psi_\theta}

The supercharges are
\eqn\defso{\CS=\oint e^{-({\varphi} + i\phi)/2}
\qquad \qquad \bar\CS = \oint.
\eqn\susyalg{
\{\CS,\bar \CS\}= \{\CS,\CS\}= \{\bar \CS,\bar\CS\} = 0.}